\documentclass[a4paper,11pt]{article}
\usepackage{jheppub} 
\usepackage{xcolor}
\usepackage{physics}
\usepackage{xspace}
\usepackage{mathrsfs}
\usepackage{soul}
\usepackage{mathtools}
\usepackage{tensor}
\usepackage{tabularx}
\usepackage{cancel}
\usepackage{booktabs}

\makeatletter
\g@addto@macro\bfseries{\boldmath}
\makeatother

\newcommand{\A}{\mathcal{A}}
\newcommand{\B}{\mathcal{B}}
\newcommand{\C}{\mathcal{C}}
\newcommand{\D}{\mathcal{D}}
\newcommand{\E}{\mathcal{E}}
\newcommand{\F}{\mathcal{F}}

\title{Gauge-invariant charges of the dual graviton}
\preprint{Imperial-TP-2024-CH-8\\ \rightline{UUITP-32/24}}

\author[a]{Chris Hull,}
\author[b]{Ulf Lindstr\"{o}m,}
\author[a]{Maxwell L. Vel\'{a}squez Cotini Hutt}

\affiliation[a]{The Blackett Laboratory, Imperial College London, Prince Consort Road, London, SW7 2AZ, UK}
\affiliation[b]{Department of Physics and Astronomy, Theoretical Physics, Uppsala University, SE-751 20, Uppsala, Sweden
and Centre for Geometry and Physics, Uppsala University, Box 480, SE-75106 Uppsala, Sweden}

\emailAdd{c.hull@imperial.ac.uk}
\emailAdd{ulf.lindstrom@physics.uu.se}
\emailAdd{m.hutt22@imperial.ac.uk} 

\abstract{The free graviton theory given by linearising Einstein's theory has a dual formulation in terms of a dual graviton field. The dual graviton theory has two gauge invariances giving rise to two conserved charges, while the ADM charges of the graviton theory become magnetic charges for the dual graviton theory. 
These charges can be ill-defined in topologically non-trivial settings and we find improvement terms that can be added to these to give gauge-invariant conserved charges.  
These gauge-invariant charges, which have local expressions in both the graviton and dual graviton formulation, give topological operators of the theory that should be considered as the generators of the genuine symmetries of the theory.
}

\begin{document}
\maketitle
\flushbottom

\section{Introduction}

Magnetic charges for a gauge theory can be found by seeking the electric charges of a dual formulation. This is a useful way of approaching magnetic charges as electric charges can be found directly from the symmetries by standard methods.
For example, the theory of a $p$-form gauge field $A$ in $d$ dimensions has a dual formulation in terms of a $(d-p-2)$-form gauge field $\tilde{A}$. The formulation in terms of $A$ has an electric charge carried by $(p-1)$-branes and a magnetic charge carried by $(d-p-3)$-branes. For the dual formulation, the charge carried by $(p-1)$-branes is a magnetic charge and that carried by the $(d-p-3)$-branes is an electric one. Thus the magnetic charge of the original theory can be found from the electric charge of the dual theory.

This approach was used for linearised gravity in~\cite{HullYetAppear}. For the theory of the free graviton $h_{\mu\nu}$ in $d$ dimensional Minkowski space, there is a conserved `electric' charge $Q[k]$ for each Killing vector $k$ of Minkowski space. 
There is a dual formulation in terms of a dual graviton $D_{\nu_1\dots\nu_{d-3}|\rho}$ \cite{Hull2000, Hull2001DualityFields}, which is a gauge field in the $[d-3,1]$ representation. (Here a $[p,q]$ tensor $T$ has a Young tableau with one column of length $p$ and one of length $q$ and we write the components as $T_{\mu_1\dots\mu_{p}|\nu_1\dots\nu_{q}}$ with a `$|$' to separate sets of antisymmetrised indices.) The field strength for this is a $[d-2,2]$ tensor $S$ and the duality with the original formulation comes from setting $S$ to be the dual of the linearised Riemann tensor constructed from $h$:
\begin{equation}\label{eq:SRstar}
    \tilde{S}(D)_{\mu\nu\alpha\beta} = R(h)_{\mu\nu\alpha\beta} \, ,
\end{equation}
where the dual of the field strength is
\begin{equation}
    \tilde{S}_{\mu\nu|\rho\sigma} = -\frac{1}{(n+1)!} \epsilon_{\mu\nu\tau_1\dots\tau_{d-2}} S\indices{^{\tau_1\dots\tau_{d-2}}_{|\rho\sigma}} \, .
\end{equation}
For $d>4$, the dual graviton theory has two gauge invariances, one with a parameter which is a $[d-3,0]$-tensor and one which is a $[d-4,1]$-tensor. The analogues of the Killing vectors $k$ are certain generalised Killing tensors, a $[d-3,0]$-tensor $\lambda$ and a $[d-4,1]$-tensor $\kappa$. These satisfy generalised Killing equations so that when they are used as gauge parameters they give invariances that leave the dual graviton field $D$ unchanged.\footnote{In particular, the differential condition on $\lambda$ requires it to be a Killing-Yano $(d-3)$-form.} These tensors were then used to construct `electric' conserved charges $Q[\kappa]$, $Q[\lambda]$ for the dual graviton theory which can then be interpreted as magnetic charges for the graviton theory. In particular, the magnetic charges for the graviton are topological in the sense that they are integrals of closed forms.

The background spacetime here is taken to be Minkowski space, possibly with some points or regions removed. In general, the gauge fields $h$ or $D$ will be defined locally in patches with the gauge fields in overlapping patches related by gauge transformations, and this structure supports certain topological or magnetic charges, which will vanish in the cases in which the gauge field is globally defined over the entire spacetime, or in which the transition functions define a trivial  structure. See \cite{Hull:2023dgp} for further discussion of the topology of such field configurations.

The charges $Q[k]$, $Q[\kappa]$, $Q[\lambda]$ for the graviton theory are all integrals of non-covariant integrands and in Ref.~\cite{Hull:2024xgo, Hull:2024mfb} manifestly covariant expressions were found which give a linear combination of these charges and a topological charge. 
Penrose, in Ref.~\cite{Penrose1982Quasi-localRelativity}, 
had considered covariant 2-form currents 
\begin{equation}\label{eq:Penrose2form}
    Y[K]_{\mu\nu} = R_{\mu\nu \alpha\beta} K^{\alpha\beta} \, ,
\end{equation}
and showed them to be conserved on-shell in the absence of sources provided that $K_{\alpha\beta}$ satisfies a certain differential equation (see equation \eqref{eq:CKY_equation}). Tensors satisfying this equation are known as conformal Killing-Yano (CKY) tensors. 
Integrating $\star Y[K]$ over a closed codimension-2 surface then gives a conserved charge for each CKY tensor which we refer to as a \emph{Penrose charge}.
These were studied in Ref.~\cite{Hull:2024xgo} and shown to give a linear combination of $Q[k]$ and a topological charge, with the Killing vector $k$ given by $k\propto\dd^\dag K$.
For the graviton theory, some of the Penrose charges can then be interpreted as giving a covariantisation of $Q[k]$. When the graviton field configuration is globally defined, the topological charge vanishes and the Penrose charges give precisely the $Q[k]$. However, when $h$ has non-trivial transition functions then the standard ADM expressions for the Noether charges $Q[k]$ are not gauge-invariant in general and should be replaced by the Penrose charges $Q[K]$. The remaining Penrose charges which are not associated with Killing vectors, are topological charges which can be viewed as magnetic charges for the graviton theory.

The Penrose charges can also be interpreted in the dual graviton theory using \eqref{eq:SRstar},
\begin{equation}
    Y[K]_{\mu\nu} = \tilde{S}_{\mu\nu\alpha\beta} (D) K^{\alpha\beta} \, .
\end{equation}
One of the aims of this paper is to study such covariant charges for the dual theory, where we will find that some can be interpreted as a covariantisation of the charges $Q[\lambda]$, while others are topological charges for the dual graviton theory.

In Ref.~\cite{Hull:2024mfb}, a covariant form was also found for $Q[\kappa]$ for the graviton theory. This involved 2-form currents which generalise \eqref{eq:Penrose2form} and are of the form
\begin{equation}\label{eq:Omega_intro}
    \Omega[V]_{\mu\nu} = R^{\alpha\beta\gamma\delta} V_{\mu\nu\alpha\beta|\gamma\delta} \, ,
\end{equation}
where $V$ is a [4,2]-tensor constructed from $\kappa$.
Then the charge defined by integrating $\star\Omega[V]$ over a codimension-2 cycle gives  the $Q[\kappa]$ plus a topological charge. The topological term can be viewed as an improvement term so that these charges can  be seen as a covariantisation of the $Q[\kappa]$ charges. In this work, we will  analyse the covariant currents related to the $\Omega[V]$ by the duality \eqref{eq:SRstar}, i.e.\ the charges in the dual graviton theory given by
\begin{equation}
    \Omega[V]_{\mu\nu} = \tilde{S}^{\alpha\beta\gamma\delta}V_{\mu\nu\alpha\beta|\gamma\delta} \, ,
\end{equation}
and discuss their relation to the $Q[\kappa]$ in the dual theory.
Gauge-invariant 2-form currents in the graviton theory have also been studied in \cite{Hinterbichler2023GravitySymmetries, Benedetti2022GeneralizedGraviton, Benedetti2023GeneralizedGravitons, BenedettiNoether, Gomez-Fayren2023CovariantRelativity, Hull:2024bcl, Hull:2024ism}.

In the graviton formulation of the theory, the charges $Q[k]$ give a linearisation of the ADM charges \cite{Abbott1982StabilityConstant}. In the dual theory the analogous charges are the $Q[\lambda]$, which are the Noether-type charges following from  invariances of the theory.
One of the goals of this work is to analyse the interplay between these charges under duality. We find that the correspondence between electric and magnetic charges in the two dual descriptions of the theory is  subtler than in $p$-form gauge theories, where electric and magnetic charges are simply interchanged under the duality.
In particular, some of the Penrose charges are realised as magnetic charges in both formulations of the theory, implying that the objects which carry these charges are intrinsically magnetic. Similarly, there are other charges which appear as electric charges in both formulations and the corresponding objects must therefore be intrinsically electrically coupled.

In section~\ref{sec:SummaryOfResults} we give a summary of the  set of covariant charges  constructed in this paper and in Refs.~\cite{Hull:2024xgo,Hull:2024mfb}.

\section{Dual formulations of linear gravity}

\subsection{The dual graviton}

In this section we give a short review of the dual graviton theory, which is an alternative formulation of the degrees of freedom of the free graviton on Minkowski space.\footnote{A different notion of duality in linearised gravity has been discussed in \cite{Nieto:1999pn, Nieto:2003de}.} For a more detailed account of both the standard and dual formulations of the theory, see \cite{HullYetAppear} and references therein.

Let us begin with the case of $d>4$ dimensions. We will return to the dual description in four dimensions in section~\ref{sec:four_dimensions_dual}.
The dual graviton is a $[d-3,1]$ tensor gauge field\footnote{We refer to a $GL(d,\mathbb{R})$ tensor transforming in the irreducible $[p,q]$ representation as a $[p,q]$ tensor. Such a tensor satisfies
\begin{equation}
    A_{\mu_1\dots\mu_p|\nu_1\dots\nu_q} = A_{[\mu_1\dots\mu_p]|\nu_1\dots\nu_q} = A_{\mu_1\dots\mu_p|[\nu_1\dots\nu_q]} \qc A_{[\mu_1\dots\mu_p|\nu_1]\nu_2\dots\nu_q} = 0 \, ,\nonumber
\end{equation}
and, in the case of $p=q$, it also satisfies $A_{\mu_1\dots\mu_p|\nu_1\dots\nu_p} = A_{\nu_1\dots\nu_p|\mu_1\dots\mu_p}$.}
$D$, with a gauge symmetry under transformations\footnote{We use square brackets to denote anti-symmetrisation with strength 1, e.g.\ $A_{[\mu\nu]} = \frac{1}{2}(A_{\mu\nu} - A_{\nu\mu})$.}
\begin{equation}\label{eq:D_gauge_transf}
    \delta D_{\mu_1\dots\mu_n|\nu} = \partial_{[\mu_1} \alpha_{\mu_2\dots\mu_n]|\nu} + \partial_\nu \beta_{\mu_1\dots\mu_n} - \partial_{[\nu} \beta_{\mu_1\dots\mu_n]} \, ,
\end{equation}
where $n\equiv d-3$. Here, $\alpha$ is an $[n-1,1]$ tensor and $\beta$ is an $n$-form.

The ``connection''
\begin{equation}\label{eq:DualConnectionDef1}
    \tilde{\Gamma}_{\mu\nu_1\dots\nu_n|\rho} = \partial_{[\mu} D_{\nu_1\dots\nu_n]|\rho} \, ,
\end{equation}
 is in the irreducible $[n+1,1]$ representation. The curvature of the dual graviton is then
\begin{equation}\label{eq:S=ddD}
    S_{\sigma_1\dots\sigma_{n+1}|\mu\nu} = \partial_{[\sigma_1} D_{\sigma_2\dots\sigma_{n+1}]|[\mu,\nu]} \, ,
\end{equation}
where the comma denotes a partial derivative, which is an   $[n+1,2]$ tensor. It satisfies the Bianchi identities
\begin{gather} 
    S_{[\sigma_1\dots\sigma_{n+1}|\mu]\nu} = 0 \, , \label{eq:S_Bianchi} \\
    \partial_{[\rho} S_{\sigma_1\dots\sigma_{n+1}]|\mu\nu} = 0 \qc S_{\sigma_1\dots\sigma_{n+1}|[\mu\nu,\rho]} = 0 \, . \label{eq:S_DifferentialBianchi}
\end{gather}

It will be convenient to define the dual of the connection
\begin{align} \label{eq:GammaTilde_Def}
    (\star \tilde{\Gamma})_{\mu\nu|\rho} = -\frac{1}{(n+1)!} \epsilon_{\mu\nu \sigma_1\dots\sigma_{n+1}} \tilde{\Gamma}\indices{^{\sigma_1 \dots \sigma_{n+1}}_{|\rho}} \, ,
\end{align}
and the dual of the field strength,
\begin{equation} \label{eq:Def.Stilde}
    \tilde{S}_{\mu\nu|\rho\sigma} = -\frac{1}{(n+1)!} \epsilon_{\mu\nu\tau_1\dots\tau_{n+1}} S\indices{^{\tau_1\dots\tau_{n+1}}_{|\rho\sigma}} \, ,
\end{equation}
which can be written
\begin{equation} \label{eq:Stile_startildeGamma}
    \tilde{S}_{\mu\nu\rho\sigma} = (\star\tilde{\Gamma})_{\mu\nu|[\rho,\sigma]} \, .
\end{equation}
In terms of $\tilde{S}$, the Bianchi identities \eqref{eq:S_Bianchi} and \eqref{eq:S_DifferentialBianchi} are
\begin{gather}
    \tilde{S}\indices{_{\mu\nu|\rho}^\nu} = 0 \, ,\label{eq:Stilde_trace=0} \\
    \partial^\mu \tilde{S}_{\mu\nu|\rho\sigma} = 0 \qc \tilde{S}_{\mu\nu|[\rho\sigma,\lambda]} = 0 \, . \label{eq:Stilde_diff_Bianchi}
\end{gather}

We define an ``Einstein tensor'' for the dual graviton,
\begin{equation}
    E_{\mu_1\dots\mu_n|\nu} = -\frac{2}{n+2} \left( S\indices{_{\mu_1\dots\mu_n \sigma|\nu}^\sigma} - \frac{n}{2} S\indices{_{\sigma\rho [\mu_1\dots\mu_{n-1}|}^{\sigma\rho}}\eta_{\mu_n]\nu} \right) \, ,
\end{equation}
which is an $[n,1]$ tensor and, as a result of \eqref{eq:S_DifferentialBianchi}, satisfies
\begin{equation} \label{eq:dual_Einstein_divergence}
    \partial^{\mu_1} E_{\mu_1\dots\mu_n|\nu} = 0\qc \partial^\nu E_{\mu_1\dots\mu_n|\nu} = 0 \, .
\end{equation}
Sources for the dual graviton are described by a `magnetic' $[n,1]$ stress tensor $U$ which must also satisfy \eqref{eq:dual_Einstein_divergence}. The sourced field equations are then \cite{Medeiros2003ExoticDuality}
\begin{equation}\label{eq:E=U}
    E_{\mu_1\dots\mu_n|\nu} = U_{\mu_1\dots\mu_n|\nu} \, .
\end{equation}
In regions where $U=0$, the field equation implies that $\tilde{S}$ satisfies
\begin{equation}\label{eq:Stilde_alg_Bianchi}
    \tilde{S}_{[\mu\nu|\rho]\sigma} \doteq 0
\end{equation}
where $\doteq$ denotes an on-shell equality. Throughout, we will be interested in quantities which are conserved only when $E=0$. This is the analogue of a Ricci-flatness in the graviton formulation of the theory. For a given field configuration of $D$, this may be satisfied at some points without the use of the field equation. Alternatively, we could impose the field equation (i.e. work on-shell) and focus on regions of spacetime where $U=0$. We will take the latter approach and impose the field equation throughout, making clear when certain equalities only hold in regions where $U=0$.

\subsection{Gravitational duality}

The standard formulation of the graviton theory is in terms of a spin-2 field $h$. The linearised {``connection''} is the [2,1] tensor\footnote{{In the linear theory, $\Gamma_{\mu\nu|\rho}$ differs from the linearisation of the Christoffel connection only by a gauge transformation, and so we will refer to it as a connection.}}
\begin{equation}
    \Gamma_{\mu\nu|\rho} = \partial_{[\mu} h_{\nu]|\rho} \, ,
\end{equation}
and the field strength is the linearised Riemann tensor,
\begin{equation}
    R\indices{_{\mu\nu}^{\rho\sigma}} = 2 \partial_{[\mu} \partial^{[\rho} h\indices{_{\nu]}^{\sigma]}} \, ,
\end{equation}
which is a [2,2] tensor. The field equation is
\begin{equation}
    G_{\mu\nu} = R_{\mu\nu} - \frac{1}{2}R\eta_{\mu\nu} = T_{\mu\nu}
\end{equation}
where $T_{\mu\nu}$ is the energy-momentum tensor.

In regions where $U=0$, the dual field strength $\tilde{S}_{\mu\nu\rho\sigma}$ satisfies \eqref{eq:Stilde_trace=0}, \eqref{eq:Stilde_diff_Bianchi}, and \eqref{eq:Stilde_alg_Bianchi} on-shell. These are precisely the same relations satisfied by the linearised Riemann tensor $R(h)_{\mu\nu\rho\sigma}$ on-shell in the standard graviton formulation of the theory in regions where $T_{\mu\nu}=0$. 
As a result, these two descriptions are dual to one another and that the duality can be imposed by the identification
\begin{equation} \label{eq:duality}
    \tilde{S}(D)_{\mu\nu\alpha\beta} = R(h)_{\mu\nu\alpha\beta} \, ,
\end{equation}
where we emphasise that the dual field strength $S$ is a function of the dual graviton $D$, whereas the linearised Riemann tensor $R$ is a function of the graviton $h$.

We can furthermore use the gauge symmetry \eqref{eq:D_gauge_transf} to choose a gauge in which the connection $\Gamma(h)$ is related to $\tilde{\Gamma}(D)$ by
\begin{equation} \label{eq:duality_gauge_choice}
    \Gamma(h)_{\alpha\beta|\mu} = -\frac{1}{2}(\star\tilde{\Gamma}(D))_{\alpha\beta|\mu} \, .
\end{equation}
In this gauge, we have
\begin{equation}
    \Gamma_{[\alpha\beta|\mu]} = 0\qc \Gamma\indices{_{\alpha\beta|}^{\beta}} = 0 \, ,
\end{equation}
where the second equation follows from the irreducibility of $\tilde{\Gamma}$; that is, $\tilde{\Gamma}_{[\mu_1\dots\mu_n|\nu]} = 0$.
Similarly,
\begin{equation}
    \tilde{\Gamma}_{[\mu_1\dots\mu_n|\nu]} = 0 \qc \tilde{\Gamma}\indices{_{\mu_1\dots\mu_n|}^{\mu_n}} = 0 \, ,
\end{equation}
where the second equation follows from the irreducibility of $\Gamma$; that is, $\Gamma_{[\mu\nu|\rho]}=0$.

\subsection{Global symmetries and charges}
\label{sec:global_symmetries}

In this subsection we construct charges for the dual graviton theory analogous to the construction of charges for the graviton theory given in \cite{Hull:2024xgo}, which was in turn a linearised version of the construction of the ADM charges given in \cite{Abbott1982StabilityConstant}.
The graviton $h_{\mu\nu}$ has gauge symmetry 
$\delta h_{\mu\nu}=2\partial_{(\mu} \xi_{\nu)}$.
For any Minkowski space Killing vector $k$ satisfying
\begin{equation}\label{eq:Killing_equation}
    \partial_{(\mu} k_{\nu)} = 0 \, ,
\end{equation}
there is a 0-form symmetry given by a gauge transformation 
with $\xi_\mu = \alpha k_\mu $ with constant 0-form parameter $\alpha$ for which the conserved Noether 1-form current is 
\begin{equation}
  j[k]_\mu =G_{\mu\nu}k^\nu
\end{equation}
where $G_{\mu\nu}$ is the linearised Einstein tensor. 
The transformations with constant $\alpha$ leave the graviton unchanged; such global symmetries were referred to as \emph{invariances} in \cite{Hull:2024xgo}.
There is then a 2-form current $J[k]_{\mu\nu}$ satisfying
\begin{equation}\label{eq:divJ=jk}
    j[k]_\mu = \partial^\nu J[k]_{\mu\nu} 
\end{equation}
that is given by
\begin{equation} \label{eq:J[k]_def}
    J[k]^{\mu\nu} = \partial_\sigma \mathcal{K}^{\mu\nu|\rho\sigma} k_\rho - \mathcal{K}^{\mu\sigma|\rho\nu} \partial_\sigma k_\rho \, ,
\end{equation}
where \begin{equation} \label{eq:DefK}
    \mathcal{K}^{\mu\nu|\rho\sigma} = -3\eta^{\mu\nu\alpha|\rho\sigma\beta} h_{\alpha\beta} \, .
\end{equation}
and
\begin{equation}
    \eta^{\alpha_1\dots\alpha_p|\beta_1\dots\beta_p} = \eta^{\alpha_1\mu_1}\cdots \eta^{\alpha_p\mu_p} \delta^{\beta_1\dots\beta_p}_{\mu_1\dots\mu_p} \, .
\end{equation}

A 2-form current $J$ satisfying $\dd^\dag J=j$ for a Noether current $j$ was referred to as a 2-form secondary Noether current in \cite{Hull:2024xgo}. Here, $J[k]$ is conserved, $\dd^\dag J[k]=0$ in regions in which $G_{\mu\nu}=0$. Using the linearised Einstein equations 
$G_{\mu\nu}=T_{\mu\nu}$ these are regions in which the energy-momentum tensor $T_{\mu\nu}$  vanishes.
There is then a Noether  charge associated with the conservation of $J[k]$,
\begin{equation}\label{eq:Q[k]}
    Q[k] = \int_\Sigma \star J[k] \, ,
\end{equation}
where $\Sigma$ is a codimension-2 cycle within a region where $T_{\mu\nu}=0$. This gives the linearised version of the ADM charges.

The dual graviton has a pair of invariances which lead to global symmetries, as we now review. Firstly, consider an $n$-form $\lambda$ satisfying
\begin{equation} \label{eq:lambda_KY}
    \partial_\mu \lambda_{\nu_1\dots\nu_n} - \partial_{[\mu}\lambda_{\nu_1\dots\nu_n]} = 0 \, .
\end{equation}
There is then a 1-form current
\begin{equation}\label{eq:j[lambda]}
    j[\lambda]_\mu = \frac{1}{n!} \lambda^{\alpha_1 \dots \alpha_n} E_{\alpha_1\dots\alpha_n|\mu} \, ,
\end{equation}
which is conserved, $\partial^\mu j[\lambda]_\mu=0$, using \eqref{eq:dual_Einstein_divergence} and \eqref{eq:lambda_KY}. Similarly, consider a $[n-1,1]$ tensor $\kappa$ satisfying
\begin{equation} \label{eq:dkappa=0}
    \partial_{[\mu}\kappa_{\nu_1\dots\nu_{n-1}]|\rho} = 0 \, .
\end{equation}
Then the 1-form current
\begin{equation}\label{eq:j[kappa]}
    j[\kappa]_\mu = \frac{1}{(n-1)!} \kappa^{\alpha_1\dots\alpha_{n-1}|\nu} E_{\mu\alpha_1\dots\alpha_{n-1}|\nu}
\end{equation}
is also conserved, $\partial^\mu j[\kappa]_\mu = 0$, as a result of \eqref{eq:dual_Einstein_divergence} and \eqref{eq:dkappa=0}. 
The tensors $\lambda$ and $\kappa$ correspond to invariances of the theory as a gauge transformation \eqref{eq:D_gauge_transf} with $\alpha = \kappa$ and $\beta = \lambda$ gives $\delta D=0$ and so leaves all field configurations invariant. These are the analogues of Killing vectors in the standard graviton theory \cite{Hull:2024xgo}.

Equation \eqref{eq:lambda_KY} is the rank-$n$ Killing-Yano equation for $\lambda$. On Minkowski space, the most general solution is
\begin{equation}\label{eq:lambda_solution}
    \lambda_{\mu_1\dots\mu_n} = \E_{\mu_1\dots\mu_n} + \F_{\mu_1\dots\mu_{n+1}} x^{\mu_{n+1}} \, ,
\end{equation}
where $\E$ is a constant $n$-form and $\F$ is a constant $(n+1)$-form.
Tensors satisfying \eqref{eq:dkappa=0} have been discussed in \cite{HullYetAppear, Hull:2024mfb} and are a type of generalised Killing tensor \cite{Howe2018SCKYT}.
It will be convenient to define the dual tensors
\begin{equation}
    \tilde{\kappa}_{\alpha\beta\gamma\delta|\mu} = \frac{1}{(n-1)!} \epsilon_{\alpha\beta\gamma\delta\sigma_1\dots\sigma_{n-1}} \kappa\indices{^{\sigma_1\dots\sigma_{n-1}}_{|\mu}} \, ,
\end{equation}
and
\begin{equation} \label{eq:LambdaDef}
    \tilde{\lambda}_{\alpha\beta\gamma} = \frac{1}{n!} \epsilon_{\alpha\beta\gamma\sigma_1\dots\sigma_n} \lambda^{\sigma_1\dots\sigma_n} \,.
\end{equation}
It follows from \eqref{eq:lambda_KY} that $\tilde{\lambda}$ is given explicitly by
\begin{equation}\label{eq:tildelambda_solution}
    \tilde{\lambda}_{\alpha\beta\gamma} = (\star \E)_{\alpha\beta\gamma} + 3(-1)^n (\star \F)_{[\alpha\beta} x_{\gamma]} \, .
\end{equation}
(In fact, $\tilde{\lambda}$ is a closed CKY 3-form; see \cite{Howe2018SCKYT, Frolov2008HigherdimensionalVariables}.)
It follows from the irreducibility of $\kappa$ and \eqref{eq:dkappa=0} that $\tilde{\kappa}$ satisfies
\begin{equation} \label{eq:kappatilde_properties}
    \tilde{\kappa}_{\alpha\beta\gamma\delta|\mu} = \tilde{\kappa}_{[\alpha\beta\gamma\delta]|\mu}\qc \partial^\alpha \tilde{\kappa}_{\alpha\beta\gamma\delta|\mu} = 0\qc \tilde{\kappa}\indices{_{\alpha\beta\gamma\delta|}^{\delta}} = 0 \, ,
\end{equation}
but $\tilde{\kappa}_{[\alpha\beta\gamma\delta|\mu]} \neq0$.
In terms of $\tilde{\lambda}$, we have, from \eqref{eq:j[lambda]},
\begin{equation} \label{eq:j[lambda]_dual_form}
    j[\lambda]_\mu = \frac{(-1)^n}{n+2} \tilde{\lambda}^{\alpha\beta\gamma} \tilde{S}_{\mu\alpha\beta\gamma}\, .
\end{equation}

Since the 1-form currents $j[\kappa]$ and $j[\lambda]$ are co-closed, they are locally co-exact and can be written as
\begin{equation}\label{eq:divJ=j}
    j[\kappa]_\mu = \partial^\nu J[\kappa]_{\mu\nu}\qc j[\lambda]_\mu = \partial^\nu J[\lambda]_{\mu\nu}
\end{equation}
for 2-forms $J[\kappa]$ and $J[\lambda]$ which are referred to as the secondary Noether currents in \cite{HullYetAppear}.
On-shell, i.e. using \eqref{eq:E=U}, in regions where $U=0$ the 1-form currents $j[\kappa]$ and $j[\lambda]$ vanish, so the secondary Noether currents $J[\kappa]$ and $J[\lambda]$ are conserved in such regions.
Hence there are charges $Q[\lambda]$ and $Q[\kappa]$ given by surface integrals over a codimension-2 cycle $\Sigma$,
\begin{equation}\label{eq:Q[lambda],Q[kappa]}
    Q[\lambda] = \int_{\Sigma} \star J[\lambda] \qc Q[\kappa] = \int_{\Sigma} \star J[\kappa] \, .
\end{equation}
These charges are then topological in the sense that they are invariant under deformations of the surface $\Sigma$ provided that $U=0$ on $\Sigma$ and the deformation does not cross sources.
The secondary Noether currents can be written \cite{HullYetAppear}
\begin{equation}\label{eq:J[kappa]_nice_def}
    J[\kappa]_{\mu\nu} = \frac{(-1)^d}{d-1} (\star\tilde{\Gamma})^{\gamma\delta|\alpha} \tilde{\kappa}_{\mu\nu\alpha\gamma|\delta} \, ,
\end{equation}
and
\begin{gather}
    J[\lambda]^{\mu\nu} = -\frac{2}{n!} \lambda_{\alpha_1\dots\alpha_n} \partial_\sigma \mathcal{L}^{\alpha_1\dots\alpha_n[\mu|\nu]\sigma} - \frac{n}{(n+1)!}\mathcal{L}^{\alpha_1\dots\alpha_n\sigma|\mu\nu} \partial_\sigma \lambda_{\alpha_1\dots\alpha_n} \, , \label{eq:J[lambda]_def}
\end{gather}
where
\begin{equation} \label{eq:DefL}
    \mathcal{L}^{\mu_1\dots\mu_n\rho|\alpha\beta} = -\eta^{\lambda\mu_1\dots\mu_n\rho|\tau_1\dots\tau_n\alpha\beta} D_{\tau_1\dots\tau_n|\lambda} \, .
\end{equation}

From the general solution \eqref{eq:lambda_solution} for $\lambda$, the $Q[\lambda]$ charges can be written
\begin{equation}\label{eq:dual_ADM_def}
    Q[\lambda] = \frac{1}{n!} \E_{\mu_1\dots\mu_n} \hat{P}^{\mu_1\dots\mu_n} + \frac{1}{(n+1)!} \F_{\mu_1 \dots \mu_{n+1}} \hat{L}^{\mu_1\dots\mu_{n+1}} \, .
\end{equation}
The charges $Q[\kappa]$ and $Q[\lambda]$ have a form analogous to that of the ADM charges in \cite{Abbott1982StabilityConstant} (see also \cite{Hull:2024xgo}). 
In particular, $\hat{P}$ is an $n$-form charge analogous to the ADM momenta, and $\hat{L}$ is an $(n+1)$-form analogous to the ADM angular momenta.\footnote{Our notation differs from \cite{HullYetAppear}, which labels the charges $Q[\lambda]$ with non-constant $\lambda$ by $\hat{J}$ instead of $\hat{L}$.}

In \cite{HullYetAppear} it was shown that the 2-forms $J[\kappa]$, which were defined in terms of $D$, can be written locally in terms of the graviton field $h$ in the duality gauge \eqref{eq:duality_gauge_choice}. The same is true for the current $J[\lambda]$ when $\lambda$ is a constant $n$-form; that is, for the $\E$-type solutions in \eqref{eq:lambda_solution}. The charges associated with these 2-forms 
{are topological charges for the graviton theory, as they are the integrals of closed forms, and 
can  be interpreted as magnetic charges of the graviton theory \cite{HullYetAppear}. }
For non-constant $\lambda$ (i.e.\ for the $\F$-type solutions in \eqref{eq:lambda_solution}) the $J[\lambda]$ cannot be written locally in terms of the graviton field, and so appear to be non-local charges for the graviton theory.
In the following, we will show that a topological term can be added to $Q[\lambda]$ for the dual graviton theory so that the resulting charge can be re-expressed in terms of the graviton field.

\subsection{Four dimensions}
\label{sec:four_dimensions_dual}

In four dimensions, the dual graviton field is a symmetric tensor gauge field $D_{\mu\nu}$ and so is of the same type as the original graviton field $h_{\mu\nu}$. In this case, the two types of gauge transformations in \eqref{eq:D_gauge_transf} degenerate and reduce to the form of a linearised diffeomorphism:
\begin{equation}
    \delta D_{\mu\nu} = 2\partial_{(\mu} \tilde{\xi}_{\nu)} \, ,
\end{equation}
for a 1-form parameter $\tilde{\xi}$.
Therefore, invariances of the four-dimensional dual graviton are given by parameters $\tilde{\xi}_\mu = \alpha \tilde{k}_\mu$, where $\tilde{k}$ is a Killing vector.\footnote{While both $k$ and $\tilde{k}$ are Killing vectors of Minkowski space, they will play a slightly different rôle in the following.} The corresponding global symmetries are then analogous to those of the graviton discussed in section~\ref{sec:global_symmetries}. There is a 2-form current
\begin{equation}
    \tilde{J}[\tilde{k}]^{\mu\nu} = \partial_\sigma \tilde{\mathcal{K}}^{\mu\nu|\rho\sigma} \tilde{k}_\rho - \tilde{\mathcal{K}}^{\mu\sigma|\rho\nu} \partial_\sigma \tilde{k}_\rho \, ,
\end{equation}
where
\begin{equation}
    \tilde{\mathcal{K}}^{\mu\nu|\rho\sigma} = -3\eta^{\mu\nu\alpha|\rho\sigma\beta} D_{\alpha\beta} \, .
\end{equation}
which is conserved, $\dd^\dag \tilde{J}[\tilde{k}] =0$, in regions where $E_{\mu\nu}=0$
where $E_{\mu\nu}$ is the Einstein tensor for the dual graviton $D_{\mu\nu}$.
On using the dual graviton field equations, $E_{\mu\nu} = U_{\mu\nu}$, these are regions where the dual energy-momentum tensor vanishes: $U_{\mu\nu}=0$. There is then a charge,
\begin{equation}
    \tilde{Q}[\tilde{k}] = \int_{\Sigma} \star \tilde{J}[\tilde{k}] \, ,
\end{equation}
where $\Sigma$ is a codimension-2 cycle contained in such a region. 

Let us compare the four-dimensional charges $\tilde{Q}[\tilde{k}]$ with the charges $Q[\lambda]$ and $Q[\kappa]$ which are present for the higher-dimensional dual graviton. We note that 
the current $J[\lambda]$ in \eqref{eq:J[lambda]_def} gives, 
for $d=4$,  the current $\tilde{J}[\tilde{k}]$. That is, the charges $Q[\lambda]$ reduce to the charges $\tilde{Q}[\tilde{k}]$ in $d=4$. 
This is in accordance with the fact that the defining equation for the KY $(d-3)$-form $\lambda$, \eqref{eq:lambda_KY}, reduces to the Killing vector equation in $d=4$.
The defining equation of the $\kappa$-tensors, \eqref{eq:dkappa=0},  implies that the $\kappa$ are constant 1-forms in $d=4$ dimensions. These are a subset of the Killing vectors $\lambda$, and the $Q[\kappa]$  become a subset of the $Q[\lambda]$.

\section{Penrose charges and their relations to Noether charges}
\label{sec:PenroseMagneticRelation}

In this section, we discuss gauge-invariant 2-form currents in the dual graviton theory and their relation to the charges $Q[\lambda]$ and $Q[\kappa]$. We begin with a brief review of the analogous construction in the graviton theory following \cite{Hull:2024xgo}, outlining the relation of these gauge-invariant currents to the ADM charges $Q[k]$. 

\subsection{Improved Penrose 2-form}

Several recent works \cite{Hinterbichler2023GravitySymmetries, Benedetti2023GeneralizedGravitons, BenedettiNoether, Gomez-Fayren2023CovariantRelativity, Hull:2024mfb, Hull:2024xgo, Hull:2024ism} have studied 2-form currents in the graviton theory that are built from contractions of the Riemann tensor and a CKY tensor. These  are conserved on-shell in regions where $T_{\mu\nu}=0$. 
Instead of the minimal currents introduced in \cite{Penrose1982Quasi-localRelativity}, we consider the improved currents
with components\footnote{This is identical in form to a linearisation of the Kastor-Traschen current \cite{Kastor2004ConservedTensors} in the case that $K$ is co-closed.}
\begin{equation}\label{eq:Y+[K]_def}
    Y_+[K]_{\mu\nu} = R_{\mu\nu\alpha\beta} K^{\alpha\beta} + 4 R\indices{^\alpha_{[\mu}} K_{\nu]\alpha} + R K_{\mu\nu} \, ,
\end{equation}
where $K$ is a conformal Killing-Yano 2-form \cite{Tachibana1969OnSpace,Kashiwada1968}; that is, it is a 2-form that satisfies
\begin{equation}\label{eq:CKY_equation}
    \partial_\rho K_{\mu\nu} = \tilde{K}_{\rho\mu\nu} + 2 \eta_{\rho[\mu} \hat{K}_{\nu]} \, ,
\end{equation}
where we use the notation $\tilde{K}=3\dd K$, $\hat{K}=\frac{1}{d-1} \dd^\dag K$, i.e.\ 
\begin{equation}
    \tilde{K}_{\rho\mu\nu} = \partial_{[\rho} K_{\mu\nu]}\qc \hat{K}_\mu = \frac{1}{d-1} \partial^\nu K_{\nu\mu} \, .
\end{equation}
There is then a conserved charge for each CKY 2-form,
\begin{equation}\label{eq:Penrose_charge}
    Q[K] = \int_\Sigma \star Y_+[K] \, ,
\end{equation}
where $\Sigma$ is a codimension-2 cycle within a region in which $T_{\mu\nu}=0$.
In \cite{Hull:2024mfb, Hull:2024xgo, Hull:2024ism}, the $Y_+[K]$ have been referred to as  improved Penrose 2-form currents, and   $Q[K]$ as   Penrose charges.

On Minkowski space, the general solution to \eqref{eq:CKY_equation} is \cite{penrose_rindler_1986,Howe2018SCKYT,Hinterbichler2023GravitySymmetries}
\begin{equation}\label{eq:CKY_solution}
    K_{\mu\nu} = \A_{\mu\nu} + \B_{[\mu} x_{\nu]} + \C_{\mu\nu\rho} x^\rho + 2x_{[\mu}\D_{\nu]\rho}x^\rho + \frac{1}{2} \D_{\mu\nu} x_\rho x^\rho \, ,
\end{equation}
where $\A$, $\B$, $\C$ and $\D$ are constant anti-symmetric tensors and $x^\mu$ are the coordinates of the background Minkowski spacetime. The divergence, $\hat{K}$, is a Killing vector,
\begin{equation}
    \hat{K}_\mu = -\frac{1}{2} \B_\mu + \D_{\mu\nu} x^\nu \, .
\end{equation}
The $\B$-type CKY 2-forms relate to constant (translational) Killing vectors whereas the $\D$-type ones relate to Lorentz  Killing vectors. The $\A$- and $\C$-type CKY 2-forms are co-closed and do not yield Killing vectors.

The main result of \cite{Hull:2024xgo} is that $Y_+[K]$ is related to the secondary Noether current $J[k]$ by
\begin{align}\label{eq:Y_+=J[k]}
    Y_+[K]^{\mu\nu} = J[k]^{\mu\nu} + \partial_\rho Z[K]^{\mu\nu\rho} \, ,
\end{align}
where
\begin{equation}\label{eq:Z[K]}
    Z[K]^{\mu\nu\rho} = 12 K^{[\mu\nu} \Gamma\indices{^{\rho\beta]}_{|\beta}} + 4h\indices{_\beta^{[\mu}} \tilde{K}^{\nu\rho\beta]}
\end{equation}
and the Killing vector $k$ is related to $K$ by
\begin{equation}\label{eq:k=Khat}
    k_\mu = 2(d-3) \hat{K}_\mu \, .
\end{equation}
Integrating \eqref{eq:Y_+=J[k]} gives a relation between the charges:
\begin{equation}\label{eq:Q[K]=Q[k]+div}
    Q[K] = Q[k] + \int_\Sigma \dd \star Z[K] \, .
\end{equation}
Since $Z[K]$ is not globally defined for general gauge field configurations $h$, the integrand of the final term, $\dd\star Z[K]$, need not be exact and so its integral over a cycle $\Sigma$ need not vanish. In the case in which $h$ is globally defined, this contribution does vanish and the Penrose charge for a $K$ that is a $\B$- or $\D$-type CKY tensor is equal to the ADM charge for Killing vector $\hat{K}$.
When $K$ is $\A$- or $\C$-type, $k=0$ and so $Q[k]=0$ and the result is only non-vanishing when $h$ is not globally defined; in this case it gives a topological charge. Several examples of this matching of charges have been studied in \cite{Hull:2024xgo}.

\subsection{Improved dual Penrose 2-form}
\label{sec:dual_Penrose_currents}

We now discuss the analogous construction in the dual graviton theory. Analogously to the $Y_+[K]$, we wish to construct conserved 2-form currents from contractions of the field strength $S$ with a $(d-2)$-form $L$. We begin with the simplest such contraction:
\begin{equation}\label{eq:Ytilde[L]}
    \tilde{Y}[L]_{\mu\nu} = \frac{2}{(n+1)!} S_{\alpha_1\dots\alpha_{n+1}|\mu\nu} L^{\alpha_1\dots\alpha_{n+1}} \, ,
\end{equation}
with a prefactor chosen for later convenience.
Demanding that this is conserved on-shell in regions where $U=0$ then gives a constraint on $L$. Namely, it is required to satisfy a higher-rank version of equation \eqref{eq:CKY_equation}, making it a CKY $(d-2)$-form. CKY $p$-forms have the property that their Hodge dual is a CKY $(d-p)$-form \cite{Howe2018SCKYT, Frolov2008HigherdimensionalVariables}. Therefore, we can set $L=\star K$ where $K$ is a CKY 2-form (i.e. is of the form \eqref{eq:CKY_solution}). We then denote the conserved 2-forms \eqref{eq:Ytilde[L]} as $\tilde{Y}[K]$. 

Much like the improved Penrose 2-form in \eqref{eq:Y+[K]_def}, there is a set of improvement terms which can be added to $\tilde{Y}[K]$ such that it remains conserved on-shell where $U=0$, but its divergence takes a nicer form off-shell.
In particular, we define the improved dual Penrose 2-form
\begin{equation} \label{eq:DualImprovedPenrose_expanded}
\begin{split}
    \tilde{Y}_+[K]_{\mu\nu} &= \frac{2}{(n+1)!} \Big( S_{\alpha_1\dots\alpha_{n+1}|\mu\nu} (\star K)^{\alpha_1\dots\alpha_{n+1}} -2(n+1) S\indices{_{\alpha_1\dots\alpha_n\sigma|[\mu}^\sigma} (\star K)\indices{^{\alpha_1\dots\alpha_n}_{\nu]}} \\
    &\qquad\qquad\qquad + \frac{n(n+1)}{2} S\indices{_{\alpha_1\dots\alpha_{n-1}\sigma\rho|}^{\sigma\rho}} (\star K)\indices{^{\alpha_1\dots\alpha_{n-1}}_{\mu\nu}} \Big) \, ,
\end{split}
\end{equation}
where $K$ is a CKY 2-form, and its Hodge dual has components
\begin{equation}
    (\star K)_{\alpha_1\dots\alpha_{n+1}} = \frac{1}{2!} \epsilon_{\alpha_1\dots\alpha_{n+1}\beta\gamma} K^{\beta\gamma} \, .
\end{equation}
$\tilde{Y}_+[K]$ is a gauge-invariant 2-form defined in terms of the dual graviton $D$. 
In terms of $\tilde{S}$, \eqref{eq:DualImprovedPenrose_expanded} can be written neatly as
\begin{align} \label{eq:Def.DualImprovedPenrose}
    \tilde{Y}_+[K]_{\mu\nu} = \tilde{S}_{\mu\nu|\rho\sigma} K^{\rho\sigma} \, .
\end{align}
Taking its divergence yields
\begin{align}
    \partial^\nu \tilde{Y}_+[K]_{\mu\nu} &= \partial^\nu \tilde{S}_{\mu\nu\rho\sigma} K^{\rho\sigma} + \tilde{S}_{\mu\nu\rho\sigma} \partial^\nu K^{\rho\sigma} \, .
\end{align}
The differential Bianchi identity \eqref{eq:Stilde_diff_Bianchi} implies that the first term vanishes. 
Then, for a CKY 2-form $K$, \eqref{eq:CKY_equation} implies
\begin{align}\label{eq:divtildeY+}
    \partial^\nu \tilde{Y}_+[K]_{\mu\nu} &= \tilde{S}_{\mu\nu\rho\sigma} \left( \tilde{K}^{\nu\rho\sigma} +2 \eta^{\nu\rho}\hat{K}^\sigma \right) = \tilde{S}_{\mu\alpha\beta\gamma} \tilde{K}^{\alpha\beta\gamma} \, ,
\end{align}
where we have used \eqref{eq:Stilde_trace=0} in the second equality.
We note that $\tilde{Y}_+[K]$ is conserved off-shell when $K$ is a closed CKY tensor, for which $\tilde{K} = 0$.\footnote{In this respect, $\tilde{Y}_+[K]$ is a generalisation of the linearised Kastor-Traschen current \cite{Kastor2004ConservedTensors} for the dual graviton.} This is the case for the $\A$- and $\B$-type CKY 2-forms in \eqref{eq:CKY_solution}.
Moreover, from \eqref{eq:divtildeY+}, in regions where $U=0$, $\tilde{Y}_+[K]$ is conserved on-shell for all $K$ using \eqref{eq:Stilde_alg_Bianchi}. Given a codimension-2 cycle $\Sigma$ on which $U=0$, we define the charges
\begin{equation}\label{eq:dual_Penrose_charge}
    \tilde{Q}[K] = \int_\Sigma \star \tilde{Y}_+[K] \, ,
\end{equation}
which we will call the dual Penrose charges. These are topological in the sense that they are unchanged under deformations of $\Sigma$ which do not cross sources.

From \eqref{eq:CKY_solution}, the general solution for $\tilde{K} \propto \dd K $ is
\begin{equation}\label{eq:Ktilde_solution}
    \tilde{K}_{\alpha\beta\gamma} = \C_{\alpha\beta\gamma} + 3\D_{[\alpha\beta}x_{\gamma]} \, ,
\end{equation}
which matches the general form of $\tilde{\lambda}$ in \eqref{eq:tildelambda_solution}. 
Note that the $\A$- and $\B$-type CKY 2-forms $K$ are closed and so do not contribute to $\tilde{K}$ in \eqref{eq:Ktilde_solution}.

Let us first consider the case of $d>4$ dimensions.
Comparing \eqref{eq:divtildeY+} with \eqref{eq:j[lambda]_dual_form} shows that $\partial^\nu \tilde{Y}_+[K]_{\mu\nu}$ is proportional to the 1-form Noether current $j[\lambda]$ off-shell, provided that we set
\begin{equation}\label{eq:tildelambda=tildeK}
    \tilde{K}_{\mu\nu\rho} = \frac{(-1)^n}{n+2} \tilde{\lambda}_{\mu\nu\rho} \, .
\end{equation}
Since any  3-form $\tilde{\lambda}$ 
of the form \eqref{eq:tildelambda_solution} agrees with  a $\tilde{K}$ of the form \eqref{eq:Ktilde_solution} for a suitable choice of 2-form $K$, no restrictions on $K$ or $\lambda$ are imposed by \eqref{eq:tildelambda=tildeK}.
Then we have
\begin{equation} \label{eq:DivDualImprovedPenrose}
    \partial^\nu \tilde{Y}_+[K]_{\mu\nu} = j[\lambda]_\mu \, .
\end{equation}
Then from \eqref{eq:DivDualImprovedPenrose} $\tilde{Y}_+[K]$ should equal the secondary Noether current $J[\lambda]$ up to a conserved term. We show in appendix~\ref{app:AppendixDualRiemann} that they are indeed related by
\begin{align}\label{eq:J[lambda]_Penrose_link}
    \tilde{Y}_+[K]^{\mu\nu} &= J[\lambda]^{\mu\nu} + \partial_\alpha \tilde{Z}[K]^{\mu\nu\alpha} \, ,
\end{align}
where 
\begin{equation}\label{eq:mho_def}
    \tilde{Z}[K]^{\mu\nu\alpha} = 3 K^{\beta[\alpha} (\star\tilde{\Gamma})\indices{^{\mu\nu]}_{|\beta}} - \frac{n}{(n+1)!} \epsilon^{\mu\nu\alpha\sigma_1\dots\sigma_n} \hat{K}^\beta D_{\sigma_1\dots\sigma_n|\beta} 
\end{equation}
is a 3-form.
Taking the divergence of \eqref{eq:J[lambda]_Penrose_link} and using \eqref{eq:divJ=j} reproduces \eqref{eq:DivDualImprovedPenrose}. 

Integrating \eqref{eq:J[lambda]_Penrose_link} over a codimension-2 cycle $\Sigma$ in a region where $U=0$ gives a relation of the charges:
\begin{equation}\label{eq:dualPenrose_ADM_Charges}
    \tilde{Q}[K] = Q[\lambda] + \int_{\Sigma} \dd \star \tilde{Z}[K] \, .
\end{equation}
The left-hand side is the dual Penrose charge associated with a CKY tensor $K$, while the first term on the right-hand side gives the $Q[\lambda]$ charge associated with the KY tensor $\lambda$ which is related to $K$ by \eqref{eq:tildelambda=tildeK}.
If $D$ is a globally defined tensor, the integrand of the final term on the right-hand side of \eqref{eq:dualPenrose_ADM_Charges} is exact, and so its integral over a cycle will vanish. Therefore, in this case, the charges $\tilde{Q}[K]$ are equal to the charges $Q[\lambda]$.
In cases where the gauge field configuration is not globally defined, this term is not exact and its integral need not vanish. In these cases, this term can be thought of as a magnetic charge for the dual graviton.

When $K$ is closed, the 3-form $\tilde{\lambda}$ related to $K$ by \eqref{eq:tildelambda=tildeK} vanishes. Therefore, the dual Penrose charges $Q[\lambda]$ are related to the dual improved Penrose charges $\tilde{Q}[K]$ for which $K$ is not closed. In summary, the $\C$- and $\D$-type CKY 2-forms in \eqref{eq:CKY_solution} are related to the $Q[\lambda]$ charges, while the $\A$- and $\B$-type CKY 2-forms give magnetic charges for the dual graviton which contribute only when $D$ is not globally defined.

In $d=4$ dimensions, we recall from section~\ref{sec:four_dimensions_dual} that the $Q[\lambda]$ charges reduce to the charges $\tilde{Q}[\tilde{k}]$ associated with Killing vectors $\tilde{k}$. The manipulations above for $d>4$ remain valid and the relation \eqref{eq:dualPenrose_ADM_Charges} becomes
\begin{equation}
    \tilde{Q}[K] = \tilde{Q}[\tilde{k}] + \int_\Sigma \dd\star \tilde{Z}[K] \, ,
\end{equation}
where $\tilde{k}$ is related to $K$ by
\begin{equation}\label{eq:ktilde=*dK}
    \tilde{K}_{\mu\nu\rho} = -\frac{1}{3} \epsilon_{\mu\nu\rho\sigma} \tilde{k}^\sigma \, .
\end{equation}
Note that this is the same as the relation \eqref{eq:tildelambda=tildeK}, with $\lambda = k$.\footnote{Previously $\lambda$ was defined to be a KY $(d-3)$-form. In $d=4$ dimensions, this is a KY 1-form, which is equivalent to a Killing vector.}

\subsection{Triviality of \texorpdfstring{$\A$}{A}- and \texorpdfstring{$\C$}{C}-type dual Penrose charges in \texorpdfstring{$d>4$}{d>4} dimensions}
\label{sec:trivial_AC}

It was shown in \cite{Benedetti2023GeneralizedGravitons, Hull:2024xgo} that in $d>4$ dimensions, the Penrose charges $Q[K]$ corresponding to $\A$- and $\C$-type CKY tensors vanish {for the graviton theory}. 
Since the dual formulation is an equivalent description of the theory, its global symmetries should match those of the graviton description. We now analyse how the charges in the dual formulation, $\tilde{Q}[K]$, vanish when $K$ is an $\A$- or $\C$-type CKY tensor in the dual formulation. This happens because the dual improved Penrose currents associated with $\A$- and $\C$-type CKY tensors are co-exact in $d>4$ dimensions.

The $\A$- and $\C$-type CKY 2-forms are co-closed. In fact, they are co-exact and can be written as the divergence of a globally-defined CKY 3-form $M$, $K=\dd^\dag M$, as we now review. 
A CKY 3-form, $M$, satisfies
\begin{equation}\label{eq:CKY_3form_equation}
    \partial_\mu M_{\alpha\beta\gamma} = \tilde{M}_{\mu\alpha\beta\gamma} + 3 \eta_{\mu[\alpha} \hat{M}_{\beta\gamma]} \, ,
\end{equation}
where
\begin{equation}
    \tilde{M}_{\mu\alpha\beta\gamma} = \partial_{[\mu} M_{\alpha\beta\gamma]} \qc \hat{M}_{\beta\gamma} = \frac{1}{d-2} \partial^\alpha M_{\alpha\beta\gamma} \, .
\end{equation}
The most general solution to this equation is of a similar form to \eqref{eq:CKY_solution} (see \cite{Hinterbichler2023GravitySymmetries} for the general form of the solution to the CKY equation of any rank on Minkowski space). CKY 3-forms on Minkowski space have the property that their divergence, $\hat{M}$, is a KY 2-form (i.e.\ a co-closed CKY 2-form) and furthermore that all KY 2-forms can be written as the divergence of a CKY 3-form \cite{Hull:2024xgo}. It follows that the $\A$- and $\C$-type CKY 2-forms can be written as $K=\hat{M}$ for a CKY 3-form $M$.

Consider now the globally-defined 3-form 
\begin{equation}
    X[M]_{\mu\nu\rho} = \tilde{S}\indices{_{[\mu\nu}^{\alpha\beta}} M_{\rho]\alpha\beta} \, .
\end{equation}
Using \eqref{eq:CKY_3form_equation}, \eqref{eq:Stilde_trace=0}, \eqref{eq:Stilde_diff_Bianchi}, and \eqref{eq:Stilde_alg_Bianchi}, we find that its divergence is
\begin{equation}\label{eq:A,C_type_trivial}
    \partial^\rho X[M]_{\mu\nu\rho} \doteq (d-4) \tilde{S}_{\mu\nu\alpha\beta} \hat{M}^{\alpha\beta} = (d-4) \tilde{Y}_+[\hat{M}]_{\mu\nu}
\end{equation}
on-shell in regions where $U=0$.
Therefore, when $d>4$, the improved dual Penrose current $\tilde{Y}_+[K]$ is co-exact when $K$ is an $\A$- or $\C$-type CKY 2-form. It follows that the dual Penrose charges associated with these CKY 2-forms vanish on-shell:
\begin{equation}
    \tilde{Q}[K] = \int_\Sigma \star \tilde{Y}_+[K] \doteq \int_\Sigma \dd\star X[M] 
    = 0 \, , 
\end{equation}
as $\partial\Sigma=0$ and $X[M]$ is a globally defined tensor field. The $\B$- and $\D$-type CKY 2-forms $K$ are not co-exact in general and so the $\B$- and $\D$-type dual Penrose charges can be non-vanishing.

In $d=4$, however, the right-hand side of \eqref{eq:A,C_type_trivial} vanishes and $\tilde{Y}_+[K]$ is not co-exact. Therefore, all four types of CKY 2-forms $K$ give non-vanishing dual Penrose charges $\tilde{Q}[K]$ in $d=4$ dimensions.

\subsection{Analysis of the dual Penrose charges}
\label{sec:BreakdownOfDualPenroseCharges}

In this section we analyse the dual Penrose charges for each type of CKY tensor to interpret them in the dual graviton theory. 

\paragraph{$\A$-type CKY tensors.}

Consider the constant $\A$-type CKY tensors, $K_{\mu\nu}=\A_{\mu\nu}$. These have $\hat{K}=0$ and $\tilde{K}=0$. 
Therefore, \eqref{eq:J[lambda]_Penrose_link} reduces to
\begin{equation}\label{eq:dualPen_Atype}
    \tilde{Y}_+[\A]^{\mu\nu} = \partial_\rho \tilde{Z}[\A]^{\mu\nu\rho}
\end{equation}
where
\begin{equation}
    \tilde{Z}[\A]^{\mu\nu\rho} = 3\A^{\beta[\rho} (\star\tilde{\Gamma})\indices{^{\mu\nu]}_{|\beta}}
\end{equation} 
from \eqref{eq:mho_def}. 
From section~\ref{sec:trivial_AC}, the $\tilde{Q}[K]$ charges vanish in $d>4$ when $K$ is $\A$-type. When $d=4$, the charge can be non-vanishing.
If $D$ is a globally defined tensor, \eqref{eq:dualPen_Atype} implies that $\tilde{Y}_+[K]$ is co-exact and the dual Penrose charges $\tilde{Q}[\A]$ vanish.
If, in $d=4$, the dual graviton $D$ is non-globally defined, then the dual Penrose charge $\tilde{Q}[\A]$ is a magnetic charge which can be non-vanishing:
\begin{equation}
    \tilde{Q}[\A] = \int_{\Sigma} \dd \star \tilde{Z}[\A] \, .
\end{equation}

\paragraph{$\B$-type CKY tensors.}

Next, consider the $\B$-type CKY tensors, $K_{\mu\nu}=\B_{[\mu}x_{\nu]}$. These have $\hat{K}_\mu = -\frac{1}{2}\B_\mu$ and $\tilde{K}_{\mu\nu\rho} = 0$, so \eqref{eq:J[lambda]_Penrose_link} becomes
\begin{equation}
    \tilde{Y}_+[\B]^{\mu\nu} = \partial_\rho \tilde{Z}[\B]^{\mu\nu\rho}
\end{equation}
where
\begin{equation}
    \tilde{Z}[\B]^{\mu\nu\rho} = \frac{3}{2} (\B^\beta x^{[\rho} - x^\beta \B^{[\rho})(\star\tilde{\Gamma})\indices{^{\mu\nu]}_{|\beta}} + \frac{n}{2(n+1)!} \epsilon^{\mu\nu\rho\sigma_1\dots\sigma_n} \B^\beta D_{\sigma_1\dots\sigma_n|\beta}
\end{equation}
so the dual Penrose charge is again a total divergence
\begin{equation}
    \tilde{Q}[\B] = \int_{\Sigma} \dd\star\tilde{Z}[\B] \, .
\end{equation}
As for the $\A$-type charges above, if $D$ is globally defined then $\tilde{Q}[\B]$ vanishes, whereas if $D$ is not globally defined then there can be a non-zero charge. Again, these are magnetic charges for the dual graviton.

\paragraph{$\C$-type CKY tensors.}

Consider now the $\C$-type CKY tensors, $K_{\mu\nu}=\C_{\mu\nu\rho}x^\rho$, which have $\hat{K}_\mu=0$ and $\tilde{K}_{\mu\nu\rho} = \C_{\mu\nu\rho}$. 
In contrast to the $\A$- and $\B$-type CKY tensors above, the $\C$-type CKY tensors correspond to a non-zero $\lambda$ through \eqref{eq:tildelambda=tildeK}. In particular, $\lambda$ is a constant $n$-form in this case.
Therefore, for $\C$-type CKY tensors $K$, \eqref{eq:J[lambda]_Penrose_link} gives
\begin{equation}\label{eq:dual_Penrose_Ctype}
    \tilde{Y}_+[K]^{\mu\nu} = J[\lambda]^{\mu\nu} + \partial_\rho \tilde{Z}[\C]^{\mu\nu\rho} \, ,
\end{equation}
where
\begin{equation}
    \tilde{Z}[\C]^{\mu\nu\rho} = 3 x_\sigma \C^{\sigma\beta[\rho}(\star\tilde{\Gamma})\indices{^{\mu\nu]}_{|\beta}} \, ,
\end{equation}
from \eqref{eq:mho_def}. 
As noted in \eqref{eq:dual_ADM_def}, the charges $Q[\lambda]$ for constant $\lambda$ constitute an $n$-form charge $\hat{P}_{\mu_1\dots\mu_n}$ in the dual graviton theory. So integrating \eqref{eq:dual_Penrose_Ctype} over a cycle $\Sigma$ yields a relation between these charges and the dual Penrose charges for the $\C$-type CKY tensors,
\begin{equation}\label{eq:dualPen_Ctype_charge}
    \tilde{Q}[\C] = -\frac{n+2}{n!} (\star \C)_{\mu_1\dots\mu_n} \hat{P}^{\mu_1\dots\mu_n} + \int_{\Sigma} \dd\star\tilde{Z}[\C] \, .
\end{equation}
Again, the final term will vanish for globally defined gauge field configurations but may contribute when $D$ is only locally defined.

From section~\ref{sec:trivial_AC}, the $\tilde{Q}[K]$ charges vanish in $d>4$ also when $K$ is $\C$-type, but can be non-vanishing in $d=4$ dimensions. Therefore, in four dimensions \eqref{eq:dualPen_Ctype_charge} implies a relation between the $\tilde{Q}[\C]$ and the charges $\hat{P}$. 
In particular, if $D$ is globally defined then the two charges are proportional.
In $d>4$ dimensions, however, $\tilde{Q}[\C]=0$ and \eqref{eq:dualPen_Ctype_charge} implies that the $\hat{P}$ charges are themselves given by a topological term $\int \dd{\star\tilde{Z}[\C]}$. We will return to this point in section~\ref{sec:electric_magnetic_relations}.

\paragraph{$\D$-type CKY tensors.}

Finally, consider the $\D$-type CKY tensors in \eqref{eq:CKY_solution}. These have $\hat{K}_\mu = \D_{\mu\nu}x^\nu$ and $\tilde{K}_{\mu\nu\rho} = 3\D_{[\mu\nu}x_{\rho]}$. That is, in this case $\tilde{K}$ is related to the KY $n$-forms $\lambda$ which are linear in the coordinates $x^\mu$ (i.e. the $\F$-type solution in \eqref{eq:lambda_solution}). In this case, \eqref{eq:J[lambda]_Penrose_link} becomes
\begin{equation}\label{eq:dual_Penrose_Dtype}
    \tilde{Y}_+[\D]^{\mu\nu} = J[\lambda] + \partial_\rho \tilde{Z}[\D]^{\mu\nu\rho} \, ,
\end{equation}
where $\tilde{Z}[\D]$ is given by \eqref{eq:mho_def} with $K$ given by the $\D$-type CKY tensor. From \eqref{eq:dual_ADM_def}, the charges $Q[\lambda]$ for $\lambda$ linear in $x^\mu$ constitute an $(n+1)$-form charge $\hat{L}^{\mu_1\dots\mu_{n+1}}$ in the dual graviton theory. 
Integrating \eqref{eq:dual_Penrose_Dtype} over a $(d-2)$-cycle $\Sigma$ yields a relation between the dual Penrose charges for the $\D$-type CKY tensors and the $\hat{L}$ charges,
\begin{equation}
    \tilde{Q}[\D] = - \frac{n+2}{(n+1)!} (\star\D)_{\mu_1\dots\mu_{n+1}} \hat{L}^{\mu_1\dots\mu_{n+1}} + \int_{\Sigma} \dd\star\tilde{Z}[\D] \, .
\end{equation}
Again, the final term will vanish when the gauge field $D$ is globally defined, but may contribute when it is not.

\subsection{Dual \texorpdfstring{$(d-3)$}{(d-3)}-form symmetries}

It has been argued in \cite{CasiniCompleteness} that higher-form symmetries should come in dual pairs and so, in particular, there should be the same number of 1-form symmetries as $(d-3)$-form symmetries. For example,  in Maxwell theory  there is one of each. This has been demonstrated in the graviton formulation of linear gravity in \cite{Benedetti2023GeneralizedGravitons, Hull:2024xgo} and we therefore expect the same result to hold in the dual formulation.
Let us briefly review how this result is seen in the graviton formulation before proceeding to the dual description.
The 1-form symmetries of the graviton are generated by the Penrose charges $Q[K]$ in \eqref{eq:Penrose_charge}. In four dimensions, there are 20 independent Penrose charges, generating a group $\mathbb{R}^{20}$ of 1-form symmetries.
As commented in section~\ref{sec:trivial_AC}, in $d>4$ dimensions the $\A$- and $\C$-type Penrose charges $Q[K]$ vanish in the graviton formulation, leaving only the $\B$- and $\D$-type charges generating a group $\mathbb{R}^{d(d+1)/2}$ of 1-form symmetries. 
In the graviton formulation, the $(d-3)$-form symmetries are generated by charges
\begin{equation}\label{eq:q[K]}
    q[K] = \int_\Sigma Y_+[K] \, ,
\end{equation}
where $\Sigma$ is a 2-cycle. In four dimensions, this charge is conserved (i.e. does not depend on deformations of $\Sigma$) for all CKY 2-forms $K$. However, they are not independent of the Penrose charges $Q[K]$ in $d=4$, and so the total number of 1-form symmetries remains as 20.
In $d>4$ dimensions, $q[K]$ is only conserved if $K$ is a closed CKY 2-form (i.e. an $\A$- or $\B$-type CKY tensor) and generate a group $\mathbb{R}^{d(d+1)/2}$ of $(d-3)$-form symmetries. 
Indeed, we see the same number of 1-form and $(d-3)$-form symmetries in all dimensions.

Let us now check the same result holds in the dual formulation of the graviton theory.
In section~\ref{sec:trivial_AC} we found that in $d=4$ all CKY 2-forms $K$ give rise to non-vanishing dual Penrose charges $\tilde{Q}[K]$, whereas in $d>4$ dimensions only the $\B$- and $\D$-type CKY tensors give non-vanishing charges. Counting the number of independent CKY tensors, this gives a 1-form symmetry group $\mathbb{R}^{20}$ in $d=4$ and $\mathbb{R}^{d(d+1)/2}$ in $d>4$ dimensions. 
Let us now consider the $(d-3)$-form symmetries. A continuous $(d-3)$-form symmetry is related to a conserved $(d-2)$-form current or, equivalently, a closed 2-form current. Therefore, let us analyse when the improved dual Penrose 2-form $\tilde{Y}_+[K]$ is closed on-shell. In regions where $U=0$, \eqref{eq:Stilde_diff_Bianchi} and \eqref{eq:Stilde_alg_Bianchi} imply that $\partial_{[\mu} \tilde{S}_{\alpha\beta]\gamma\delta}=0$ on-shell, so we have
\begin{equation}\label{eq:Ytilde_closed}
    \partial_{[\rho} \tilde{Y}_+[K]_{\mu\nu]} = \tilde{S}\indices{_{[\mu\nu}^{\alpha\beta}} \partial_{\rho]} K_{\alpha\beta} = \tilde{S}\indices{_{[\mu\nu}^{\alpha\beta}} \tilde{K}_{\rho]\alpha\beta} -2 \tilde{S}_{\beta[\rho\mu\nu]}\hat{K}^\beta = \tilde{S}\indices{_{[\mu\nu}^{\alpha\beta}} \tilde{K}_{\rho]\alpha\beta} \, ,
\end{equation}
where we have used the CKY equation \eqref{eq:CKY_equation} in the second equality and \eqref{eq:Stilde_alg_Bianchi} in the final equality.

In $d>4$, \eqref{eq:Ytilde_closed} implies that $\tilde{Y}_+[K]$ is closed when $\tilde{K}=0$, which is the case for the $\A$- and $\B$-type solutions in \eqref{eq:CKY_solution}. 
Therefore, when $K$ is an $\A$- or $\B$-type CKY 2-form, we define the conserved charges
\begin{equation}\label{eq:tildeq[K]}
    \tilde{q}[K] = \int_{\Sigma} \tilde{Y}_+[K] \, ,
\end{equation}
where $\Sigma$ is a 2-cycle contained in a region where $U=0$. These charges generate a group $\mathbb{R}^{d(d+1)/2}$ of $(d-3)$-form symmetries of the dual graviton theory, matching the number of 1-form symmetries found above.

When $d=4$, however, \eqref{eq:Ytilde_closed} implies that $\tilde{Y}_+[K]$ is closed for all CKY 2-forms $K$. This can be seen by dualising,
\begin{equation}
    \frac{1}{3!} \epsilon^{\rho\mu\nu\tau_1\dots\tau_n} \partial_\rho \tilde{Y}_+[K]_{\mu\nu} = \frac{1}{3} S^{\rho\tau_1\dots\tau_n|\mu\nu} \tilde{K}_{\rho\mu\nu} \, .
\end{equation}
In $d=4$, we have $n=d-3=1$ and so the result vanishes on-shell from the algebraic Bianchi identity \eqref{eq:S_Bianchi}.
This would seem to imply that there are a further 20 one-form symmetries in $d=4$ generated by the $\tilde{q}[K]$. However, the $\tilde{q}[K]$ are not independent of the $\tilde{Q}[K]$ and they generate the same 20 one-form symmetries. This can be seen by dualising $\tilde{S}$ and using standard properties of the Levi-Civita symbol:
\begin{equation}
\begin{split}
    \tilde{Y}_+[K]_{\mu\nu} &= \tilde{S}_{\mu\nu\alpha\beta} K^{\alpha\beta} = \frac{1}{4} \epsilon_{\mu\nu\rho\sigma} \epsilon^{\alpha\beta\kappa\lambda} S\indices{^{\rho\sigma}_{\alpha\beta}} (\star K)_{\kappa\lambda} = -6 S\indices{^{\rho\sigma}_{[\mu\nu}} (\star K)_{\rho\sigma]} \\
    & = -S_{\mu\nu\rho\sigma} (\star K)^{\rho\sigma} = - \frac{1}{2} \epsilon_{\mu\nu\alpha\beta} \tilde{S}^{\alpha\beta\rho\sigma} (\star K)_{\rho\sigma} = - \star \tilde{Y}_+[\star K]_{\mu\nu} \, .
\end{split}
\end{equation}
This equation relates the dual improved Penrose current associated with a CKY 2-form $K$ to the Hodge dual of the dual improved Penrose current associated with $\star K$ in four dimensions. However, the since the Hodge dual of a CKY 2-form is itself a CKY 2-form \cite{Howe2018SCKYT, Frolov2008HigherdimensionalVariables},
\begin{equation}
    \tilde{q}[K] = \int_S \tilde{Y}_+[K] = - \int_S \star \tilde{Y}_+[\star K] = \tilde{Q}[-\star K] \, .
\end{equation}
and so are equal to the dual Penrose charges $\tilde{Q}[K']$ for another CKY tensor $K'=-\star K$. Therefore, there are only 20 independent charges in $d=4$ dimensions, which can be parametrised by the $\tilde{Q}[K]$.

\subsection{Covariant currents for the \texorpdfstring{$\kappa$}{kappa}-charges}

Consider the case of $d>4$ dimensions. Then there are charges $Q[\lambda]$ and $Q[\kappa]$ associated with invariances of the dual graviton. We have seen that the dual Penrose charges give a covariant expression for the $Q[\lambda]$ in the previous subsections. We now consider the $Q[\kappa]$ and a covariant expression for them.

As remarked in section~\ref{sec:global_symmetries}, the 2-form currents $J[\kappa]$, defined in terms of $D$ in \eqref{eq:J[kappa]_nice_def}, can be written locally in terms of $h$ in the duality gauge \eqref{eq:duality_gauge_choice}. 
In \cite{Hull:2024mfb}, assuming this choice of gauge, the covariant 2-form $\Omega[V]$ in \eqref{eq:Omega_intro} was built from a contraction of the Riemann tensor with a [4,2]-tensor $V$ such that
\begin{equation}\label{eq:Omega=J[kappa]}
    \Omega[V]_{\mu\nu} = J[\kappa]_{\mu\nu} + \partial^\rho \Phi_{\mu\nu\rho}
\end{equation}
provided that $V$ satisfies
\begin{equation}\label{eq:divV=kappa}
    \partial^\mu V_{\mu\nu\alpha\beta|\gamma\delta} = \frac{(-1)^d}{d-1} \tilde{\kappa}_{\nu\alpha\beta[\gamma|\delta]} \, .
\end{equation}
In \eqref{eq:Omega=J[kappa]}, we have defined the 3-form
\begin{equation}\label{eq:Phi_def}
    \Phi_{\mu\nu\rho} = 2 \Gamma^{\gamma\delta|\beta} V_{\mu\nu\alpha\beta|\gamma\delta} \, .
\end{equation}
One simple solution of the condition \eqref{eq:divV=kappa} is
\begin{equation}\label{eq:V_example}
    V_{\mu\nu\alpha\beta|\gamma\delta} = \frac{(-1)^d}{d-1} \tilde{\kappa}_{\mu\nu\alpha\beta|[\gamma} x_{\delta]} \, ,
\end{equation}
where $x^\mu$ are the coordinates of the background Minkowski spacetime.\footnote{If the background spacetime has non-trivial topology, e.g. toroidal directions, then the components of $V$ given in \eqref{eq:V_example} are not necessarily well-defined. In some cases, it is still possible to find explicit solutions to \eqref{eq:divV=kappa} \cite{Hull:2024mfb}.}
Integrating \eqref{eq:Omega=J[kappa]} over a codimension-2 cycle, we have a relation between the charge associated with the covariant current $\Omega[V]$, the $Q[\kappa]$ charge, and a topological charge,
\begin{equation}\label{eq:Q[V]=Q[kappa]}
    \mathbb{Q}[V] \equiv \int_\Sigma \star \Omega[V] = Q[\kappa] + \int_\Sigma \dd\star\Phi \, ,
\end{equation}
such that the $\mathbb{Q}[V]$ can be seen as a covariantisation of the $Q[\kappa]$ defined only in terms of the Riemann tensor.

We now describe the analogous construction for the dual graviton formulation.
Consider the 2-forms
\begin{equation}\label{eq:tildeOmega_def}
    \tilde{\Omega}[V]_{\mu\nu} = \tilde{S}^{\gamma\delta\alpha\beta} V_{\mu\nu\alpha\beta|\gamma\delta} \, ,
\end{equation}
related to $\Omega[V]$ in \eqref{eq:Omega_intro} by the duality \eqref{eq:duality}.
Then we have
\begin{equation}\label{eq:tildeOmega=J[kappa]}
    \tilde{\Omega}[V]_{\mu\nu} = \partial^\delta (\star \tilde{\Gamma})^{\alpha\beta|\gamma} V_{\mu\nu\alpha\beta|\gamma\delta} = J[\kappa]_{\mu\nu} + \partial^\rho \tilde{\Phi}_{\mu\nu\rho} \, ,
\end{equation}
where $\tilde{\Phi}$ is a 3-form with components
\begin{equation}\label{eq:tildePhi_def}
    \tilde{\Phi}_{\mu\nu\rho} = (\star\tilde{\Gamma})^{\gamma\delta|\alpha} V_{\mu\nu\alpha\rho|\gamma\delta} \, ,
\end{equation}
which is related to $\Phi$ by the gravitational duality provided we choose the gauge \eqref{eq:duality_gauge_choice}.
In reaching \eqref{eq:tildeOmega=J[kappa]} we have used \eqref{eq:Stile_startildeGamma} in the first equality and \eqref{eq:J[kappa]_nice_def} in the second. It then follows from the on-shell conservation of $J[\kappa]$ in the dual graviton theory that $\tilde{\Omega}[V]_{\mu\nu}$ is conserved on-shell in regions where $U=0$. Integrating over a codimension-2 cycle $\Sigma$ contained in such a region, the charge defined by integrating $\star\tilde{\Omega}[V]$ is related to $Q[\kappa]$ and a topological charge:
\begin{equation}\label{eq:tildeOmega_charge_Q[kappa]}
    \tilde{\mathbb{Q}}[V] \equiv \int_\Sigma \star \tilde{\Omega}[V] = Q[\kappa] + \int_\Sigma \dd \star \tilde{\Phi} \, .
\end{equation}
From \eqref{eq:tildePhi_def}, $\tilde{\Phi}$ is globally defined when $D$ is globally defined. In cases where it is global, the integrand of final term in \eqref{eq:tildeOmega_charge_Q[kappa]} is exact and so its integral vanishes. We then find that the gauge-invariant charge $\tilde{\mathbb{Q}}[V]$ gives precisely the charge $Q[\kappa]$ which was defined in terms of the non-gauge-invariant current $J[\kappa]$. For non-globally defined field configurations, the final term will contribute in general. We can then see the current $\tilde{\Omega}[V]$ as the covariantisation of the $J[\kappa]$ current.

As mentioned in the previous subsection, it has been suggested that 1-form and $(d-3)$-form symmetries come in dual pairs. The existence of non-trivial $\mathbb{Q}[V]$ and $\tilde{\mathbb{Q}}[V]$ charges depends on some non-trivial topology of the background spacetime (e.g. compact dimensions, or excision of some points or regions) \cite{Hull:2024mfb}. These charges generate 1-form symmetries, and it would be interesting to understand if there are additional $(d-3)$-form symmetries dual to them on such non-trivial backgrounds.

\section{Relations of gravitational electric and magnetic charges}
\label{sec:electric_magnetic_relations}

Consider a region without sources in either formulation, so $T_{\mu\nu}=0$ and $U_{\mu_1\dots\mu_n|\nu}=0$. In such a region, the theory can be formulated in terms of either the graviton $h$ or the dual graviton $D$. The formulations are related via the duality \eqref{eq:duality}. We will furthermore assume the choice of gauge in \eqref{eq:duality_gauge_choice}. 

The invariances of the graviton are given by Killing vectors $k$ in \eqref{eq:Killing_equation} and lead to conserved charges $Q[k]$ given in \eqref{eq:Q[k]}. The invariances of the dual graviton are given by KY $n$-forms $\lambda$ in \eqref{eq:lambda_KY} and generalised Killing tensors $\kappa$ satisfying \eqref{eq:dkappa=0}, which lead to conserved charges $Q[\lambda]$ and $Q[\kappa]$ in \eqref{eq:Q[lambda],Q[kappa]}.

In the graviton description, there is a gauge-invariant conserved 2-form $Y_+[K]$ given by \eqref{eq:Y+[K]_def} for each CKY 2-form of Minkowski space, $K$. These lead to the conserved Penrose charges $Q[K]$ in \eqref{eq:Penrose_charge}. In the dual graviton description, there is a gauge-invariant conserved 2-form $\tilde{Y}_+[K]$ given by \eqref{eq:Def.DualImprovedPenrose} for each $K$, which lead to conserved dual Penrose charges $\tilde{Q}[K]$ in \eqref{eq:dual_Penrose_charge}. 

In the graviton description, the $Q[k]$ are related to the Penrose charges $Q[K]$ by \eqref{eq:Q[K]=Q[k]+div}, while in the dual graviton description the $Q[\lambda]$ are related to the dual Penrose charges $\tilde{Q}[K]$ by \eqref{eq:dualPenrose_ADM_Charges}. Furthermore, the $Q[\kappa]$ charges of the dual graviton are related to the covariant charges $\tilde{\mathbb{Q}}[V]$ by \eqref{eq:tildeOmega_charge_Q[kappa]}. Furthermore, the analogues of the $Q[\kappa]$ in the graviton theory, which are defined only in the gauge \eqref{eq:duality_gauge_choice}, are related to the covariant charges $\mathbb{Q}[V]$ by \eqref{eq:Q[V]=Q[kappa]}.

Note that on-shell in the graviton theory we have $R_{\mu\nu}=0$, and so the improved Penrose 2-form \eqref{eq:Y+[K]_def} becomes
\begin{equation}
    Y_+[K] \doteq R_{\mu\nu\alpha\beta} K^{\alpha\beta} \, ,
\end{equation}
where $\doteq$ denotes an on-shell equality.
Then under the duality \eqref{eq:duality}, we have
\begin{equation}
    Y_+[K] \doteq \tilde{Y}_+[K]
\end{equation}
on-shell, which implies that the Penrose and dual Penrose charges are equal on-shell in regions without sources:
\begin{equation}\label{eq:Q[K]=Qtilde[K]}
    Q[K] \doteq \tilde{Q}[K] \, .
\end{equation}
In addition, the 2-forms $\Omega[V]$ and $\tilde{\Omega}[V]$, which are covariantisations of the $J[\kappa]$ current in the graviton and dual graviton formulation respectively, are also related by the duality \eqref{eq:duality}, such that
\begin{equation}
    \mathbb{Q}[V] = \tilde{\mathbb{Q}}[V] \, .
\end{equation}

\subsection{Four dimensions}
\label{sec:4d_relations_with_duality}

In $d=4$ dimensions, there are 20 CKY 2-forms $K$ and the Penrose and dual Penrose charges corresponding to all of them are non-vanishing. The results of section~\ref{sec:PenroseMagneticRelation} give a relation between these charges and the charges $Q[k]$ and $\tilde{Q}[\tilde{k}]$ associated with invariances of the graviton and dual graviton respectively. Here, $k$ and $\tilde{k}$ are Killing vectors related to $K$ by \eqref{eq:k=Khat} and \eqref{eq:ktilde=*dK} respectively. 

In particular, the $\A$-type CKY 2-forms have $k=0$ and $\tilde{k}=0$. Therefore, \eqref{eq:Q[K]=Q[k]+div} and \eqref{eq:dualPenrose_ADM_Charges} imply that the $\A$-type Penrose charges are `magnetic' in both the graviton and dual graviton formulations; that is, the $\A$-type currents $Y_+[\A]$ and $\tilde{Y}_+[\A]$ are both identically conserved, and the corresponding charge is an integral of a total divergence. As such, the charge is non-vanishing only when $h$ and $D$ are not globally defined. 
This behaviour is not seen in standard $p$-form gauge theories, where electric and magnetic charges are interchanged under the duality.
Having charges which are magnetic in both formulations is novel to gauge theories with fields in more general Lorentz representations. Commonly, e.g. in Maxwell theory, the objects which carry magnetic charges in one formulation of the theory are solitonic, but can be described locally in the dual formulation where they carry electric charge. The fact that the $\A$-type Penrose charges are magnetic in both formulations implies that the objects which carry these charges are intrinsically solitonic. In \cite{Hinterbichler2023GravitySymmetries} it was found that one solution carrying the $\A$-type charge in the graviton description is a linearisation of the C-metric.

The $\B$-type CKY 2-forms have non-zero and constant $k=-\B/2$, but $\tilde{k}=0$. That is, they relate to the ADM momenta, but not to any of the $\tilde{Q}[\tilde{k}]$. Therefore, \eqref{eq:Q[K]=Q[k]+div} implies that the $\B$-type Penrose charges give the sum of the ADM momenta and a magnetic charge $\int\dd\star Z[K]$ in the graviton formulation, while \eqref{eq:dualPenrose_ADM_Charges} implies that they are purely magnetic in the dual graviton formulation, i.e. of the form $\int\dd\star \tilde{Z}[K]$. Consider a field configuration where $h$ is globally defined but $D$ is not, then $\int\dd\star Z[K]=0$ and so equating \eqref{eq:Q[K]=Q[k]+div} and \eqref{eq:dualPenrose_ADM_Charges} gives
\begin{equation}\label{eq:B_type_duality}
    Q[k] = \int_\Sigma \dd\star\tilde{Z}[K] \, ,
\end{equation}
where the left-hand side is the ADM momentum associated with a constant Killing vector $k$, and the right-hand side is a magnetic charge for the dual graviton $D$. This is the expected behaviour of an electro-magnetic duality, where an electric charge in one formulation is manifest as a magnetic charge in the dual formulation. This is juxtaposed with the $\A$-type charges discussed above, where the charge is magnetic in both formulations. 

The $\C$-type charges have the opposite behaviour since a $\C$-type CKY tensor corresponds to $k=0$ but a constant non-zero $\tilde{k}=-3 \star\C$ in four dimensions. Therefore \eqref{eq:Q[K]=Q[k]+div} implies that the $\C$-type Penrose charges are magnetic in the graviton formulation but \eqref{eq:dualPenrose_ADM_Charges} implies that they give the sum of the $\tilde{Q}[\tilde{k}]$ and a magnetic charge $\int \dd\star \tilde{Z}[K]$ in the dual graviton formulation. Consider a field configuration for which $D$ is globally defined but $h$ is not, then $\int \dd\star\tilde{Z}[K]=0$, so equating \eqref{eq:Q[K]=Q[k]+div} and \eqref{eq:dualPenrose_ADM_Charges} gives
\begin{equation}
    \tilde{Q}[\tilde{k}] = \int_\Sigma \dd\star Z[K] \, .
\end{equation}
Therefore, the charges $\tilde{Q}[\tilde{k}]$ corresponding to constant $\tilde{k}$ are electric charges in the dual graviton formulation, but magnetic in the graviton description.

Finally, consider the $\D$-type Penrose charges. The $\D$-type CKY 2-forms have $k_\mu = \D_{\mu\nu} x^\nu$ and $\tilde{k}_{\mu} = - \epsilon_{\mu\nu\rho\sigma}\D^{\nu\rho} x^{\sigma}$ and so correspond to the Lorentz boost Killing vectors. Then \eqref{eq:Q[K]=Q[k]+div} implies that the $\D$-type Penrose charges give the sum of the ADM angular momenta and a magnetic charge in the $h$ formulation. On the other hand, \eqref{eq:dualPenrose_ADM_Charges} implies that the $\D$-type Penrose charges give the sum of the $\tilde{Q}[\tilde{k}]$ and a magnetic charge in the $D$ formulation. We remark that the $\D$-type Penrose charges are electric in both formulations of the theory; that is, $Y_+[\D]$ and $\tilde{Y}_+[\D]$ are only conserved on-shell in both formulations. Similarly to the $\A$-type charges, this behaviour is not observed in Maxwell theory where electric and magnetic charges are interchanged under duality. This implies that the objects carrying the $\D$-type charges are intrinsically `electric' in both formulations. In \cite{Hinterbichler2023GravitySymmetries} it was shown that a linearisation of the Kerr metric gives a four-dimensional solution carrying the $\D$-type Penrose charge.

\subsection{\texorpdfstring{$d>4$}{d>4} dimensions}

In $d>4$ dimensions, we found in section~\ref{sec:trivial_AC} that the $\A$- and $\C$-type Penrose charges vanish on-shell when they are defined on a cycle $\Sigma$ in a region where $U=0$. This leaves the $\B$- and $\D$-type charges. These are manifest in the graviton and dual graviton descriptions in the same way as in $d=4$. That is, the $\B$-type Penrose charges are electric and relate to the ADM momenta in the graviton formulation, but are magnetic in the dual graviton formulation. The $\D$-type Penrose charges are again electric in both formulations, relating to the ADM angular momenta in the graviton theory and the $Q[\lambda]$ (with non-constant $\lambda$) in the dual graviton theory.

It is interesting to consider the $\C$-type Penrose charges, which we know vanish on-shell in $d>4$ dimensions. $\C$-type CKY 2-forms have $k=0$ but constant non-zero $\tilde{\lambda} = (-1)^n (n+2) \C$. Equation \eqref{eq:dualPenrose_ADM_Charges} relates the $\C$-type Penrose charges to the sum of the $Q[\lambda]$ and a magnetic charge $\int\dd\star\tilde{Z}[K]$.
As the $\C$-type Penrose charges vanish, this implies that
\begin{equation}
    Q[\lambda] \doteq -\int_\Sigma \dd\star\tilde{Z}[K] \, .
\end{equation}
That is, the $Q[\lambda]$ charges associated with constant $\lambda$ can be written as the integral of a total divergence in $d>4$ dimensions in the dual graviton formulation, and so will vanish when $D$ is globally defined. This is in stark contrast to their behaviour in $d=4$, where they cannot be written in this form and can be non-vanishing even when $D$ is globally defined.

\subsection{Examples of dualities between Penrose charges}

In this subsection we discuss several examples of solutions which carry the Penrose charges associated with a given CKY tensor. This has been studied in detail in \cite{Hinterbichler2023GravitySymmetries, Hull:2024xgo} in the graviton formulation. Here, we discuss these charged solutions both in the graviton and dual graviton theories such that the duality of their Penrose charges can be made manifest.

As discussed in subsection~\ref{sec:4d_relations_with_duality}, in four dimensions there is an interesting duality between the $\B$- and $\C$-type Penrose charges. In particular, when $h$ is globally defined but $D$ is not, the $\B$-type Penrose charges are electric in the graviton description and magnetic in the dual description as in \eqref{eq:B_type_duality}. The $\C$-type charges show the opposite behaviour.

Consider the four-dimensional linearised Schwarzschild solution for the graviton. This has Cartesian components:
\begin{equation}
    h_{tt} = \frac{M}{r}\qc h_{ij} = \frac{M x_i x_j}{r^2}
\end{equation}
where $M$ is a mass parameter and $r^2 = x^i x_i$. The non-vanishing components of the linearised Riemann tensor are
\begin{equation}
\begin{split}\label{eq:h_schw_Riemann}
    R\indices{_{ij}^{kl}} &= 2M \left( \frac{3}{r^5} x^{[k}x_{[i} \delta_{j]}^{l]} - \frac{1}{r^3} \delta_{[i}^{[k} \delta_{j]}^{l]} \right) \\
    R\indices{^{it}_{jt}} &= M \left( - \frac{3 x^i x_j}{2r^5} + \frac{\delta^i_j}{2r^3} \right)
\end{split}
\end{equation}
One can verify that this solves the vacuum field equations $G_{\mu\nu}=0$. This solution carries the Penrose charge corresponding to a $\B$-type CKY tensor $K_{\mu\nu} = \B_{[\mu}x_{\nu]}$ where $\B_\mu = (\B_0, 0,0,0)$ with $\B_0$ a constant \cite{Hinterbichler2023GravitySymmetries, Hull:2024xgo}. This $h$ configuration is globally defined and one can verify \eqref{eq:Q[K]=Q[k]+div} explicitly:
\begin{equation}\label{eq:Schw_Btype_charge}
    Q[K] = Q[k] = -2\pi M \B_0 \, .
\end{equation}
That is, the $\B$-type Penrose charge reproduces the ADM mass of the solution. Here the surface on which the charges are defined has been chosen as a 2-sphere of fixed radius centred on the origin in $\mathbb{R}^3$.

We now ask what the corresponding solution for the dual graviton $D$ is, whose linearised curvature $S$ is related to $R$ in \eqref{eq:h_schw_Riemann} by \eqref{eq:duality}. The non-zero components of $S$ must be
\begin{equation}\label{eq:D_Taub_Riemann}
    S_{tijk} = \frac{M}{2} \left( -\frac{3}{r^5} x_i \epsilon_{jkl} x^l + \frac{1}{r^3} \epsilon_{ijk} \right)
\end{equation}
This is satisfied by the linearisation of the Lorentzian Taub-NUT solution \cite{Taub, NUT} in which $D$ has components
\begin{equation}
    D_{it} = 2A_i
\end{equation}
where $A_i$ is a 1-form connection on $\mathbb{R}^3\setminus \{ 0 \}$ with field strength $F=\dd{A}=-\frac{M}{2}\text{Vol}_{S^2}$ where $\text{Vol}_{S^2}$ is the volume form on the unit 2-sphere. In components this can be written
\begin{equation}
    F_{ij} = \epsilon_{ijk} \partial_k V
\end{equation}
where
\begin{equation}
    V = -\frac{M}{r}
\end{equation}
One can verify that the only non-zero components of $S$ are $S_{tijk} = \frac{1}{2} \partial_i F_{jk}$ which reproduces \eqref{eq:D_Taub_Riemann}. In the dual graviton description, the $\B$-type Penrose charge is magnetic and is given by
\begin{equation}
    Q[K] = \int_{S^2} \dd\star\tilde{Z}[K] = \B_0 \int_{S^2} F = -2\pi M \B_0
\end{equation}
which matches \eqref{eq:Schw_Btype_charge}, as was guaranteed. It is apparent that in the $D$ formulation, this charge would vanish identically if $A$ was a globally defined 1-form as then $F$ would be exact. This is, therefore, a clear example where the $\B$-type Penrose charges are electric in the $h$ description and magnetic in the corresponding $D$ description of the solution. Furthermore, it emphasises that Lorentzian Taub-NUT can be seen as carrying the dual charge of the ADM mass.

Similarly, if we had taken the Lorentzian Taub-NUT solution for $h$, we would find a non-vanishing $\C$-type Penrose charge as the gauge field is not globally defined. The corresponding description in terms of $D$ is the Schwarzschild spacetime, where the $\C$-type Penrose charge is electric. The manipulations in this case are identical to those discussed above, with $h$ and $D$ interchanged.

\section{Summary of the relationships between charges}
\label{sec:SummaryOfResults}

In this section we provide a brief summary of the results found here and in \cite{Hull:2024xgo, Hull:2024mfb}. This is intended to clarify the relationships between the different charges in different subcases.
For each type of CKY 2-form $K$ in \eqref{eq:CKY_solution}, there are Penrose charges $Q[K]$ which, in regions without sources, can be written in terms of either $h$ or $D$. Below, we give the relation between the Penrose charges $Q[K]$ and the charges $Q[k]$ and $Q[\lambda]$ (or $\tilde{Q}[\tilde{k}]$ in $d=4$) in the graviton formulation and dual graviton formulation respectively. We also review the relation of the $\mathbb{Q}[V]$ to the $Q[\kappa]$ in the dual formulation.

In each case, the Penrose charge $Q[K]$ is gauge-invariant and it is related to a topological charge, or to a Noether charge plus a topological charge. Here, we call a charge topological if it is the integral of a closed form, and will vanish unless the form represents a non-trivial cohomology class. We write the topological charges in the form $\int\dd\star Z$ for some $Z$ and so these will vanish if $Z$ is globally-defined. In the cases  in which the Penrose charge is the sum of a Noether charge and a topological charge, the Noether and topological charges are not separately gauge-invariant in general, and so the total derivative term can be thought of as an improvement added to the Noether charge to render it gauge-invariant. If the corresponding gauge field is globally-defined, the topological term is trivial and the Penrose charge is precisely the usual Noether charge.

\subsection{Penrose charges}

\subsubsection{Four dimensions}

In $d=4$ dimensions, there are 20 1-form symmetries generated by the Penrose charges $Q[K]$. Recall that there are four different types of CKY tensors (see \eqref{eq:CKY_solution}). In table~\ref{tab:4d_breakdown} we show the relations between the Penrose charges corresponding to each type of CKY tensor and the Noether charges $Q[k]$ and $\tilde{Q}[\tilde{k}]$ which are defined in the graviton and dual graviton formulations of the theory respectively. 

\begin{table}[ht]
    \centering
    \begin{tabular}{c l l}
        \hline\\[-2.5ex]
        \textbf{CKY 2-form, $K$} & \textbf{Graviton formulation} & \textbf{Dual graviton formulation} \\ [0.5ex] 
        \hline\hline \\[-2ex]
        $\A$ & $Q[\A]=\int\dd\star Z[\A]$ & $Q[\A]=\int\dd\star \tilde{Z}[\A] $ \\ [1ex] 
        $\B$ & $Q[\B]=Q[k] + \int\dd\star Z[\B]$ & $Q[\B]=\int\dd\star\tilde{Z}[\B]$ \\ [1ex] 
        $\C$ & $Q[\C]=\int\dd\star Z[\C]$ & $Q[\C]=\tilde{Q}[\tilde{k}] + \int\dd\star \tilde{Z}[\C]$ \\ [1ex] 
        $\D$ & $Q[\D]=Q[k] + \int\dd\star Z[\D]$ & $Q[\D]=\tilde{Q}[\tilde{k}] + \int\dd\star \tilde{Z}[\D]$\\[1ex]
        \hline
    \end{tabular}
    \caption{Relationships between the Penrose charges $Q[K]$ and the charges $Q[k]$ in the graviton formulation and $\tilde{Q}[\tilde{k}]$ in the dual graviton formulation. The relation is different for each  type of CKY tensor (defined by \eqref{eq:CKY_solution}). The 3-forms $Z[K]$ and $\tilde{Z}[K]$ are defined in \eqref{eq:Z[K]} and \eqref{eq:mho_def} respectively. The Killing vectors $k$ and $\tilde{k}$ are related to $K$ by \eqref{eq:k=Khat} and \eqref{eq:ktilde=*dK} respectively.
    }
    \label{tab:4d_breakdown}
\end{table}

In the graviton formulation, the situation is the following.
For $\B$- and $\D$-type CKY 2-forms $K$, $Q[K]$ gives the ADM charge $Q[k]$ for the corresponding Killing vector $k\propto\dd^\dag K$ defined by \eqref{eq:k=Khat} plus a topological term.
If $h$ is globally defined, the topological term vanishes so that 
the Penrose charge equals the ADM charge. If $h$ is not globally defined, neither $Q[k]$ nor the topological term is separately gauge-invariant, so the Penrose charge gives an `improvement' of the standard ADM expression to yield a gauge-invariant charge.
For the $\A$- and $\C$-type CKY 2-forms $K$, $\dd^\dag K=0 $ and there is no associated  Killing vector, so that the Penrose charge $Q[K]$ is  a topological charge, which is only non-vanishing if $h$ is not globally defined.

In the dual  formulation in terms of a dual graviton $D_{\mu\nu}$, the relations are slightly different.
For $\C$- and $\D$-type CKY 2-forms $K$, there is a Killing vector $\tilde{k}$ defined by \eqref{eq:ktilde=*dK}. For these CKY tensors, $Q[K]$ gives the charge $\tilde{Q}[\tilde{k}]$ for the Killing vector $\tilde{k} \propto \star\dd K$ plus a topological charge which improves the standard $\tilde{Q}[\tilde{k}]$ charge so that the sum is gauge-invariant. 
When the dual graviton is globally defined, the topological charge vanishes and $Q[K]=\tilde{Q}[\tilde{k}]$.
For the $\A$- and $\B$-type CKY 2-forms $K$, $\star\dd K=0$ and $Q[K]$ gives a topological charge, which is only non-vanishing if $D$ is not globally defined.

\subsubsection{\texorpdfstring{$d>4$}{d>4} dimensions}

\begin{table}[ht]
    \centering
    \begin{tabular}{c l l}
        \hline\\[-2.5ex]
        \textbf{CKY 2-form, $K$} & \textbf{Graviton formulation} & \textbf{Dual graviton formulation} \\ [0.5ex] 
        \hline\hline \\[-2ex] 
        $\B$ & $Q[\B]=Q[k] + \int\dd\star Z[\B]$ & $Q[\B]=\int\dd\star\tilde{Z}[\B]$ \\ [1ex] 
        $\D$ & $Q[\D]=Q[k] + \int\dd\star Z[\D]$ & $Q[\D]=Q[\lambda] + \int\dd\star \tilde{Z}[\D]$\\[1ex]
        \hline
    \end{tabular}
    \caption{Relationships between the Penrose charges $Q[K]$ and the charges $Q[k]$ in the graviton formulation and $Q[\lambda]$ in the dual graviton formulation in dimensions $d>4$. The relation is different for each type of CKY tensor (defined by \eqref{eq:CKY_solution}). The 3-forms $Z[K]$ and $\tilde{Z}[K]$ are defined in \eqref{eq:Z[K]} and \eqref{eq:mho_def} respectively. The Killing vector $k$ and KY $(d-3)$-form $\lambda$ are related to $K$ by \eqref{eq:k=Khat} and \eqref{eq:tildelambda=tildeK} respectively. The Penrose charges associated with $\A$- and $\C$-type CKY tensors vanish on-shell in $d>4$ dimensions: $Q[\A] = Q[\C] = 0$.
    }
    \label{tab:d>4_breakdown}
\end{table}

In higher dimensions, the charges $\tilde{Q}[\tilde{k}]$ are replaced by the $Q[\lambda]$ associated with KY $(d-3)$-forms $\lambda$ related to $K$ by \eqref{eq:tildelambda=tildeK}.
Note that $k$ (defined in \eqref{eq:k=Khat}) remains a Killing vector. The salient difference between $d=4$ and $d>4$ is that in higher dimensions the $\A$- and $\C$-type Penrose charges vanish: $Q[\A]=Q[\C]=0$ (see discussion in section~\ref{sec:trivial_AC}). 
The relationship of the remaining $\B$- and $\D$-type charges to the Noether charges $Q[k]$ and $Q[\lambda]$ are summarised in table~\ref{tab:d>4_breakdown}.

The vanishing of $Q[\A]$ and $Q[\C]$ in $d>4$ has particularly interesting consequences for the $\C$-type charges, where in the dual theory we find that \eqref{eq:dualPenrose_ADM_Charges} implies $Q[\lambda] = -\int\dd\star\tilde{Z}[\C]$. Therefore, the $Q[\lambda]$ charges associated with constant $\lambda$ -- which are related to particular invariances of the dual graviton involving a constant KY $(d-3)$-form -- become topological, in contrast to their nature in four dimensions.
The $\B$- and $\D$-type Penrose charges again each give a standard ADM charge plus a topological charge in the graviton formulation, providing a gauge-invariant charge that can be viewed as an improvement of the ADM charge when $h$ is not globally defined.

\subsection{\texorpdfstring{$\mathbb{Q}[V]$}{Q[V]} charges}

In $d>4$, for each Killing tensor $\kappa$ satisfying \eqref{eq:dkappa=0}, there is a tensor $V$ related to $\kappa$ by \eqref{eq:divV=kappa} which can be used to define a conserved current $\Omega[V]$ given in \eqref{eq:Omega_intro}.
This can then be integrated to give the charge $\mathbb{Q}[V]$ given in \eqref{eq:Q[V]=Q[kappa]}.
In the  dual graviton formulation, this is a Noether charge plus a topological charge:
\begin{equation}\label{eq:QV=Qkappa_dual}
    \mathbb{Q}[V] = Q[\kappa] + \int\dd\star \tilde{\Phi} \, ,
\end{equation}
where $Q[\kappa]$ is the Noether charge \eqref{eq:Q[lambda],Q[kappa]} and $\tilde{\Phi}$ is given by \eqref{eq:tildePhi_def}.
If the dual graviton is globally defined, $\mathbb{Q}[V]$ is precisely the Noether charge $Q[\kappa]$, while in the general case it provides a gauge-invariant completion of the Noether charge.
These charges can be rewritten in terms of the graviton $h$ in a particular gauge \cite{HullYetAppear}. For the graviton theory, both $Q[\kappa]$ and $\mathbb{Q}[V] $ are topological charges.

\subsection{\texorpdfstring{$(d-3)$-form}{(d-3)-form} symmetries}

There is also a set of gauge-invariant charges $q[K]$ defined in \eqref{eq:q[K]} in the graviton formulation and \eqref{eq:tildeq[K]} in the dual formulation. These are conserved when $K$ is a closed CKY 2-form (i.e. an $\A$- or $\B$-type CKY 2-form in \eqref{eq:CKY_solution}). They have support on a 2-cycle and so generate $(d-3)$-form symmetries of the theory. In $d=4$ dimensions these charges are not independent of the Penrose charges, whereas they are independent in $d>4$. Given that the $\A$- and $\C$-type Penrose charges are trivial in $d>4$ dimensions (see section~\ref{sec:trivial_AC}), there is the same number of non-trivial Penrose charges $Q[K]$ as dual charges $q[K]$.

\section{Conclusion and outlook}

We have constructed gauge-invariant 2-form conserved currents $\tilde{Y}_+[K]$ and $\tilde{\Omega}[V]$ (defined in \eqref{eq:DualImprovedPenrose_expanded} and \eqref{eq:tildeOmega_def} respectively) in the dual graviton theory. 
For certain tensors $K$ and $V$, the Hodge duals of these currents can be integrated over $(d-2)$-cycles to give charges $\tilde{Q}[K]$ and $\tilde{\mathbb{Q}}[V]$ respectively, which generate 1-form symmetries of the theory. Furthermore, for certain $K$, the $\tilde{Y}_+[K]$ can be integrated over 2-cycles to give charges $\tilde{q}[K]$ generating $(d-3)$-form symmetries.
The dual graviton field $D$ provides an alternative formulation of a free spin-2 field propagating in a Minkowski spacetime to the usual one in terms of  $h_{\mu \nu}$.
The two formulations are related by gravitational duality \cite{Hull2000, Hull2001DualityFields}, under which the charges $\tilde{Q}[K]$, $\tilde{\mathbb{Q}}[V]$ and $\tilde{q}[K]$ directly translate into the charges $Q[K]$, $\mathbb{Q}[V]$ and $q[K]$ (defined in \eqref{eq:Penrose_charge}, \eqref{eq:Q[V]=Q[kappa]} and \eqref{eq:q[K]} respectively) which are written in terms of the graviton field $h$. 
That is, the 1-form and $(d-3)$-form global symmetries which these charges generate match exactly between the two formulations of the theory.\footnote{It has been shown in \cite{Hull:2024ism} that several pairs of 1-form symmetries of the graviton theory have mixed 't Hooft anomalies. These must also be present in the dual formulation. We leave the confirmation of this to future work.} 

The charges $Q[K]$ of the graviton theory are of particular interest due to their close link to the ADM charges $Q[k]$ \cite{Hull:2024xgo}. In particular, from \eqref{eq:Q[K]=Q[k]+div}, they differ only by a topological charge. Similarly, we have shown in this work that the $\tilde{Q}[K]$ charges of the dual graviton and the charges $Q[\lambda]$, as defined in \cite{HullYetAppear}, differ only by a topological charge; see \eqref{eq:dualPenrose_ADM_Charges}. Moreover, the topological charge which enters in \eqref{eq:Q[K]=Q[k]+div} is a function of $h$ and cannot, in general, be written locally in terms of $D$. Similarly, the topological charge in \eqref{eq:dualPenrose_ADM_Charges} cannot always be written locally in terms of $h$.
This means that charges that arise naturally in one formulation of the theory become non-local in the dual formulation: there are conserved currents that can be written locally in terms of $h$ but not in terms of $D$, and vice-versa.
For example, as discussed in \cite{HullYetAppear}, the ADM angular momenta $Q[k]$ associated with non-constant Killing vectors cannot be written in terms of $D$ and the $Q[\lambda]$ associated with non-constant Killing-Yano tensors $\lambda$ cannot be written in terms of $h$. In addition, the $Q[k]$ and $Q[\lambda]$ charges are not gauge-invariant in topologically non-trivial situations, as the integrand can change by a closed form which is not exact \cite{Hull:2024xgo}.
The gauge-invariant charges $Q[K]$ and $\tilde{Q}[K]$, however, are true conserved quantities of the graviton theory. In certain circumstances (restricting the topology of the field configurations), the ADM charges $Q[k]$ and the $Q[\lambda]$ charges will also be gauge-invariant.

Given the importance of the charges $Q[K]$ and $\tilde{Q}[K]$, another objective of this work has been to investigate their properties under gravitational duality. We find that the map between electric and magnetic charges under duality is considerably more interesting for a spin-2 field than for standard $p$-form gauge fields. In particular, in the graviton description the subset of these charges corresponding to $\B$- and $\D$-type CKY tensors $K$ (see \eqref{eq:CKY_solution}) are electric charges in the sense that they are conserved on-shell only in the absence of sources. In contrast, the $\A$- and $\C$-type CKY tensors give rise to identically conserved currents and, therefore, to magnetic charges. In the dual formulation of the theory, the $\C$- and $\D$-type charges are electric, while the $\A$- and $\B$-type charges are magnetic. That is, the $\A$-type charges are magnetic in both descriptions, whereas the $\D$-type charges are electric in both. This is in  contrast to the behaviour of similar charges in $p$-form gauge theories, where electric and magnetic charges are exchanged under duality.

In this paper we have discussed currents and charges related to second rank CKY tensors. 
It is natural to seek generalisations of our discussion to
higher rank covariant currents such as 
\begin{equation}
    R\indices{_{[\mu\nu}^{\alpha\beta}} K_{\rho]\alpha\beta} \qc R\indices{_{[\mu\nu}^{\alpha\beta}} K_{\rho\sigma]\alpha\beta} \qc \dots
\end{equation}
with $K$ a higher rank CKY tensor. 
However, the corresponding charges vanish, as may be shown via the following short argument.
 
Consider the rank-$p$ current $J[K]_{\mu_1\dots\mu_p}=R\indices{_{[\mu_1\mu_2}^{\alpha\beta}}K_{\mu_3\dots\mu_p]\alpha\beta}$ where $2<p<d-2$ and $K$ is a $p$-form. Demanding that this current is conserved restricts $K$ to be a KY $p$-form (i.e.\  a co-closed CKY $p$-form). On Minkowski space, such tensors are co-exact and can be written $K=\dd^\dag M$ where $M$ is a rank-$(p+1)$ CKY tensor. Inserting this into $J[K]$ then implies that it is also co-exact: $J[K] \propto \dd^\dag J'[M]$ where $J'[M]_{\mu_1\dots\mu_{p+1}} = R\indices{_{[\mu_1\mu_2}^{\alpha\beta}}M_{\mu_3\dots\mu_{p+1}]\alpha\beta}$. Since $J'[M]$ is a function of the curvature $R$, it is globally-defined. It then follows that the charges $\int_\Sigma \star J[K]$ vanish by Stokes' theorem when $\Sigma$ is a codimension-$p$ cycle.
See also \cite{Benedetti2023GeneralizedGravitons}.

While the free graviton theory has a dual formulation in terms of the dual graviton, there is no such duality for the non-linear Einstein theory, and indeed there is not a local covariant interacting theory of the dual graviton \cite{Bekaert:2002uh}.
Nonetheless, the dual graviton theory has proved useful in constructing magnetic charges for the free graviton theory, and one can aim to extend  these to magnetic charges for the non-linear theory. We will discuss such extensions elsewhere.

\paragraph{Acknowledgements.} CH is supported by the STFC Consolidated Grants ST/T000791/1 and ST/X000575/1. 
UL gratefully acknowledges a Leverhulme Visiting Professorship to Imperial College as well as the hospitality of the theory group at Imperial.
MVCH is supported by a President’s Scholarship from Imperial College London.

\appendix

\section{Relation between \texorpdfstring{$\tilde{Y}_+[K]$}{\tilde{Y}[K]} and \texorpdfstring{$J[\lambda]$}{J[lambda]}}
\label{app:AppendixDualRiemann}

In this appendix we derive the relation given in \eqref{eq:J[lambda]_Penrose_link} between the dual improved Penrose current $\tilde{Y}_+[K]$ and the secondary Noether current $J[\lambda]$ when the theory is formulated in terms of the dual graviton $D$.
Consider the improved dual Penrose 2-form defined in \eqref{eq:Def.DualImprovedPenrose}. From the definition of $\tilde{S}_{\mu\nu\rho\sigma}$ in \eqref{eq:Stile_startildeGamma}, we have
\begin{align}
    \tilde{Y}_+[K]^{\mu\nu} &= \tilde{S}^{\mu\nu\alpha\beta} K_{\alpha\beta} \nonumber \\
    &= -K_{\alpha\beta} \partial^\alpha (\star\tilde{\Gamma})^{\mu\nu|\beta} \nonumber\\
    &= \partial_\alpha \left( K\indices{_\beta^\alpha} (\star\tilde{\Gamma})^{\mu\nu|\beta} \right) - (\star\tilde{\Gamma})^{\mu\nu|\beta} \partial_\alpha K\indices{_\beta^\alpha} \nonumber\\
    &= \partial_\alpha \left( 3 K\indices{_\beta^{[\alpha}} (\star\tilde{\Gamma})^{\mu\nu]|\beta} - 2 K\indices{_\beta^{[\mu}} (\star\tilde{\Gamma})^{\nu]\alpha|\beta} \right) - (\star\tilde{\Gamma})^{\mu\nu|\beta} \partial_\alpha K\indices{_\beta^\alpha} \nonumber\\
    &= \partial_\alpha \Pi^{\mu\nu\alpha} - 2 \partial^\alpha K^{\beta[\mu} (\star\tilde{\Gamma})\indices{^{\nu]}_{\alpha|\beta}} - 2 K\indices{_\beta^{[\mu}} \partial_\alpha (\star\tilde{\Gamma})^{\nu]\alpha|\beta} - (\star\tilde{\Gamma})^{\mu\nu|\beta} \partial_\alpha K\indices{_\beta^\alpha} \, , \label{eq:Ytilde_working0}
\end{align}
where we have integrated by parts and rearranged terms to absorb a total divergence into a 3-form $\Pi$ with components
\begin{equation}
    \Pi^{\mu\nu\alpha} = 3 K^{\beta[\alpha} (\star\tilde{\Gamma})\indices{^{\mu\nu]}_{|\beta}} \, .
\end{equation}
We now note that $S_{[\alpha_1\dots\alpha_{n+1}|\beta]\gamma}=0$ implies
\begin{equation} \label{eq:StildeRicci}
    0 = \eta^{\mu\alpha} \tilde{S}_{\mu\nu\alpha\beta} = \frac{1}{2} \left( \partial^\alpha (\star\tilde{\Gamma})_{\alpha\nu|\beta} - \partial_\beta (\star\tilde{\Gamma})\indices{^\alpha_{\nu|\alpha}} \right) \, .
\end{equation}
When the theory is formulated in terms of $D$, we have
\begin{equation}
    \tilde{\Gamma}_{[\alpha_1\dots\alpha_{n+1}|\beta]} = \partial_{[\alpha_1} D_{\alpha_2\dots\alpha_{n+1}|\beta]} = 0 \, ,
\end{equation}
which is equivalent to 
\begin{equation}\label{eq:dualconnectionTraceless}
    (\star\tilde{\Gamma})\indices{^{\nu\alpha}_{|\alpha}}=0 \, .
\end{equation}
Therefore, \eqref{eq:StildeRicci} implies that
\begin{equation}
    \partial^\alpha (\star\tilde{\Gamma})_{\alpha\nu|\beta} = 0 \, ,
\end{equation}
so the third term on the right hand side of the final equality in \eqref{eq:Ytilde_working0} vanishes. This then leaves
\begin{align}
    \tilde{Y}_+[K]^{\mu\nu} &= \partial_\alpha \Pi^{\mu\nu\alpha} - 2 \partial^\alpha K^{\beta[\mu} (\star\tilde{\Gamma})\indices{^{\nu]}_{\alpha|\beta}} - (\star\tilde{\Gamma})^{\mu\nu|\beta} \partial_\alpha K\indices{_\beta^\alpha} \nonumber \\
    &= \partial_\alpha \Pi^{\mu\nu\alpha} +(\dim-1) \hat{K}^\beta (\star\tilde{\Gamma})\indices{^{\mu\nu}_{|\beta}} - 2\left( \tilde{K}^{\alpha\beta[\mu} + \eta^{\alpha\beta}\hat{K}^{[\mu} - \hat{K}^\beta \eta^{\alpha[\mu} \right) (\star\tilde{\Gamma})\indices{^{\nu]}_{\alpha|\beta}} \nonumber \\
    &= \partial_\alpha \Pi^{\mu\nu\alpha} + (\dim-1) \hat{K}^\beta (\star\tilde{\Gamma})\indices{^{\mu\nu}_{|\beta}} - 2 \tilde{K}^{\alpha\beta[\mu} (\star\tilde{\Gamma})\indices{^{\nu]}_{\alpha|\beta}} \nonumber \\
    &\qquad -2 \left( \hat{K}^{[\mu} (\star\tilde{\Gamma})\indices{^{\nu]\alpha}_{|\alpha}} + \hat{K}^\beta (\star\tilde{\Gamma})\indices{^{\mu\nu}_{|\beta}} \right) \nonumber \\
    &= \partial_\alpha \Pi^{\mu\nu\alpha} - 2 \tilde{K}^{\alpha\beta[\mu} (\star\tilde{\Gamma})\indices{^{\nu]}_{\alpha|\beta}} +(\dim-3) \hat{K}^\beta (\star\tilde{\Gamma})\indices{^{\mu\nu}_{|\beta}} \, , \label{eq:Ytilde_working}
\end{align}
where we have used the CKY equation \eqref{eq:CKY_equation} and \eqref{eq:dualconnectionTraceless}.

We will now demonstrate that the final two terms of the previous equation give $J[\lambda]^{\mu\nu}$ up to a total divergence, where $\tilde{\lambda}_{\mu\nu\rho} = \tilde{K}_{\mu\nu\rho}$.
Firstly, consider the term in \eqref{eq:Ytilde_working} involving $\tilde{K}$. This can be written
\begin{align}
    \tilde{K}^{\alpha\beta[\mu} (\star\tilde{\Gamma})\indices{^{\nu]}_{\alpha|\beta}} &= \frac{1}{n!(n+1)!} \epsilon\indices{^{\gamma_1\dots\gamma_n\beta[\mu}_\alpha} \epsilon^{\nu]\sigma_1\dots\sigma_{n+1}\alpha} (\star\tilde{K})_{\gamma_1\dots\gamma_n} \tilde{\Gamma}_{\sigma_1\dots\sigma_{n+1}|\beta} \nonumber \\
    &= (-1)^{n+1} \frac{n+2}{n!} \eta^{\beta\gamma_1\dots\gamma_n[\mu|\nu]\kappa\sigma_1\dots\sigma_n} \partial_\kappa D_{\sigma_1\dots\sigma_n|\beta} (\star\tilde{K})_{\gamma_1\dots\gamma_n} \, ,
\end{align}
which, from \eqref{eq:DefL}, can be written
\begin{equation}
    \tilde{K}^{\alpha\beta[\mu} (\star\tilde{\Gamma})\indices{^{\nu]}_{\alpha|\beta}} = (-1)^{n+1} \frac{n+2}{2n!} (\star\tilde{K})_{\gamma_1\dots\gamma_n} \left( \partial_\kappa \mathcal{L}^{\gamma_1\dots\gamma_n\nu|\mu\kappa} - \partial_\kappa \mathcal{L}^{\gamma_1\dots\gamma_n\mu|\nu\kappa} \right) \, . \label{eq:dualPen_lambdatermResult}
\end{equation}
Consider now the term involving $\hat{K}$ in \eqref{eq:Ytilde_working}. Using the definition of $\tilde{\Gamma}$ in \eqref{eq:GammaTilde_Def},
\begin{align}
    \hat{K}^\beta (\star\tilde{\Gamma})\indices{^{\mu\nu}_{|\beta}} &= -\frac{1}{(n+1)!} \epsilon^{\mu\nu\alpha\sigma_1\dots\sigma_n} \hat{K}^\beta \tilde{\Gamma}_{\alpha\sigma_1\dots\sigma_n|\beta} \nonumber \\
    &= \partial_\alpha \Sigma^{\mu\nu\alpha} + \frac{1}{(n+1)!} \epsilon^{\nu\nu\alpha\sigma_1\dots\sigma_n} \partial_\alpha \hat{K}_\beta D\indices{_{\sigma_1\dots\sigma_n}^{|\beta}} \, , \label{eq:k.Gammatilde_working}
\end{align}
where
\begin{equation}
    \Sigma^{\mu\nu\alpha} = -\frac{1}{(n+1)!} \epsilon^{\mu\nu\alpha\sigma_1\dots\sigma_n} \hat{K}^\beta D_{\sigma_1\dots\sigma_n|\beta} \, .
\end{equation}
It is a property of CKY tensors $K$ that $\hat{K}$ is non-locally related to $\tilde{K}$ by \cite[Appendix A]{Hull:2024xgo}
\begin{equation}
    \partial_\alpha \hat{K}_\beta = -\frac{1}{n+1} \partial^\kappa \tilde{K}_{\kappa\alpha\beta} \, .
\end{equation}
Substituting this into \eqref{eq:k.Gammatilde_working} gives
\begin{align}
    \hat{K}^\beta (\star\tilde{\Gamma})\indices{^{\mu\nu}_{|\beta}} &= \partial_\alpha \Sigma^{\mu\nu\alpha} - \frac{1}{(n+1)!n!} \left( \frac{1}{n+1} \right) \epsilon_{\kappa\alpha\beta\gamma_1\dots\gamma_n} \epsilon^{\mu\nu\alpha\sigma_1\dots\sigma_n} D\indices{_{\sigma_1\dots\sigma_n}^{|\beta}} \partial^\kappa (\star\tilde{K})^{\gamma_1\dots\gamma_n} \nonumber \\
    &= \partial_\alpha \Sigma^{\mu\nu\alpha} + (-1)^{n} \frac{n+2}{(n+1)!} \eta^{\beta\gamma_1\dots\gamma_n\kappa|\sigma_1\dots\sigma_n\mu\nu} D_{\sigma_1\dots\sigma_n|\beta} \partial_\kappa (\star\tilde{K})_{\gamma_1\dots\gamma_n} \nonumber \\
    &= \partial_\alpha \Sigma^{\mu\nu\alpha} + (-1)^{n+1} \frac{n+2}{(n+1)!} \mathcal{L}^{\gamma_1\dots\gamma_n\kappa|\mu\nu} \partial_\kappa (\star\tilde{K})_{\gamma_1\dots\gamma_n} \, . \label{eq:dualPen_ktermResult}
\end{align}
Inserting the results of \eqref{eq:dualPen_lambdatermResult} and \eqref{eq:dualPen_ktermResult} in \eqref{eq:Ytilde_working} gives
\begin{align}
\begin{split}
    \tilde{Y}_+[K]^{\mu\nu} &= \partial_\alpha \Pi^{\mu\nu\alpha} + n \partial_\alpha \Sigma^{\mu\nu\alpha} \\
    & \quad + (-1)^n (n+2) \bigg[ \frac{1}{n!} (\star\tilde{K})_{\gamma_1\dots\gamma_n} \left( \partial_\kappa \mathcal{L}^{\gamma_1\dots\gamma_n\nu|\mu\kappa} - \partial_\kappa \mathcal{L}^{\gamma_1\dots\gamma_n\mu|\nu\kappa} \right) \\
    &\quad - \frac{n}{(n+1)!} \mathcal{L}^{\gamma_1\dots\gamma_n\kappa|\mu\nu} \partial_\kappa (\star\tilde{K})_{\gamma_1\dots\gamma_n} \bigg] \, .
\end{split}
\end{align}
Finally, comparing with \eqref{eq:J[lambda]_def} and defining
\begin{equation}
    \tilde{Z}[K]^{\mu\nu\alpha} = \Pi^{\mu\nu\alpha} + n \Sigma^{\mu\nu\alpha} 
\end{equation}
yields \eqref{eq:J[lambda]_Penrose_link}, where $\lambda$ and $K$ are related by \eqref{eq:tildelambda=tildeK}.

\bibliographystyle{JHEP}
\bibliography{references}

\end{document}